\documentclass[twocolumn,prx,hyperref,footinbib,superscriptaddress]{revtex4-1}

\usepackage{bbm,amssymb}
\usepackage{times}
\usepackage{graphicx}
\usepackage{dcolumn}
\usepackage{bm}
\usepackage{float}
\usepackage{mathrsfs}
\usepackage{color}
\usepackage{hyperref}

\usepackage{amsmath}

\usepackage{tabularx}
\usepackage{multirow}
\usepackage{booktabs}
\usepackage{makecell}

\newcommand{\ket}[1]{\left|#1\right>}
\newcommand{\bra}[1]{\left< #1 \right|}
\def\braket#1#2{\left\langle#1\middle|#2\right\rangle}

\begin{document}

\title{Spin polarization through Intersystem Crossing in the silicon vacancy of silicon carbide}

\author{Wenzheng Dong}
\email{dongwz@vt.edu}
\affiliation{ Department of Physics, Virginia Tech, Blacksburg, Virginia 24061, USA}
\author{M. W. Doherty}
\affiliation{Laser Physics Centre, Research School of Physics and Engineering, Australian National University,
Australian Capital Territory 2601, Australia}
\author{Sophia  E.  Economou}
\email{economou@vt.edu}
\affiliation{ Department of Physics, Virginia Tech, Blacksburg, Virginia 24061, USA}

\begin{abstract}
Silicon carbide (SiC)-based defects are promising for quantum communications, quantum information processing, and for the next generation of quantum sensors, as they feature long coherence times, frequencies near the telecom, and optical and microwave transitions. For such applications, the efficient initialization of the spin state is necessary. We develop a theoretical description of the spin polarization process by using the  intersystem crossing of the silicon vacancy defect, which is enabled by a combination of optical driving, spin-orbit coupling, and interaction with vibrational modes. By using distinct optical drives, we analyze two spin polarization channels. Interestingly, we find that different spin projections of the ground state manifold can be polarized. This work helps to understand initialization and readout of the silicon vacancy and explains some existing experiments with the silicon vacancy center in SiC. 
\end{abstract}

\maketitle

\section{Introduction}

Color centers in silicon carbide (SiC) have been of interest over the last several years as candidate platforms alternative to the NV center in diamond  for quantum information and sensing applications \cite{WeberPNAS2010,AtatureNatureMat2018,AwschalomNature2018,RadulaskiNanoLetters2017,CastellettoNatureMat2013,SiminPRX2016}. SiC is attractive due to the following properties: it has a large band gap to host deep defects \cite{SormanPRB2000} and benefits from mature fabrication techniques \cite{WidmannNatureMat2014}; it is  CMOS-compatible \cite{FuchsSciRep2013}, and it is cost-effective compared to diamond. The two most studied defects in SiC to date are the divacancy (a missing pair of neighboring Si and C atoms) \cite{KoehlNature2011,ChristleNatMat2014,AbramPRL2015,ZargalehPRB2016} and the monovacancy (a missing Si atom) \cite{JanzenPhyB2009,HainJAP2014,BracherNanoLet2015,DefoPRB2018}. Both of these vacancy centers have promising features for quantum information applications, such as long spin coherence times, even at room temperature, and both optical and microwave transitions for control \cite{KoehlNature2011,WidmannNatureMat2014}.

Like the NV center in diamond, the divacancy in SiC has six active electrons associated with it, the same total spin and a similar electronic structure. As a result, prior investigations of the NV center in diamond \cite{MazeNature2008,DohertyNJP2011} can be used to understand, at least qualitatively, the electronic structure and dynamics of the SiC divacancy. On the other hand, the Si monovacancy (henceforth referred to as V$_{\text{Si}}$) has five active electrons, leading to a half-integer total spin ($S=\frac{3}{2}$ in the ground state) and a distinct electronic structure. This high-spin character of V$_{\text{Si}}$ can provide additional capabilities of interest in applications. For example, V$_{\text{Si}}$ has been used for vector magnetometry \cite{LeePRB2015,SiminPRAP2015,NiethammerPRAP2016} and all-optical magnetometry \cite{SiminPRX2016}. In addition, this defect has been shown to feature a few different transitions for potential use in spin-photon interfaces \cite{EconomouNanoTech2016,NagyPRAP2018}.

A previous work by one of us \cite{SoykalPRB2016} found the symmetry-adapted multi-particle states of $\text{V}_{\text{Si}}$ using group theory and DFT. Going beyond the electronic structure and understanding the physics under optical drive and the microscopic mechanisms of the resulting spin polarization (optical pumping) is crucial, both for applications and for a deeper understanding of the defect. Such an analysis is currently lacking for V$_{\text{Si}}$.

In this paper we address this issue and present a detailed theoretical analysis of the intersystem crossing mechanism and the dynamics of $\text{V}_{\text{Si}}$ under optical drive. Our work examines the interplay of the physical mechanisms responsible for the generation of spin polarization, namely spin-orbit coupling (SOC) and coupling between the defect electronic states and vibrational modes, and reveals which paths among the many allowed transitions can yield spin polarization. We show that for a thorough description of this process, additional levels, not included in Ref. \cite{SoykalPRB2016}, need to be taken into account. Through numerical simulations of the optical polarization process and comparison to experiment, we can deduce typical values of the intersystem crossing rates. We find that initialization to both the $|S_z|=3/2$ and the $|S_z|=1/2$ states can occur, depending on the excited state manifold driven by the laser and the relative relaxation rates among the doublets. Our work provides a microscopic counterpart to phenomenological models that have been used to explain spin polarization experiments in $\text{V}_{\text{Si}}$ \cite{FuchsNatCom2015}. 

The paper is structured as follows. In Section II we give a brief introduction to the $C_{3v}$ point group, based on which the many body wave functions are obtained. In Section III, we introduce the concept of intersystem crossing (ISC) and the terms in the Hamiltonian that contribute to ISC in $\text{V}_{\text{Si}}$. In Section IV, we demonstrate two optically-driven spin polarization protocols from two distinct channels corresponding to two different excited state manifolds. We simulate numerically the dynamics using a Lindblad equation and show that spin polarization can be obtained efficiently within the ground quartets.

\section{Overview of   ${C}_{{3v}}$  symmetry in  ${V}_{{Si}}$ }

\begin{figure}
\centering
\includegraphics[width=.9\linewidth]{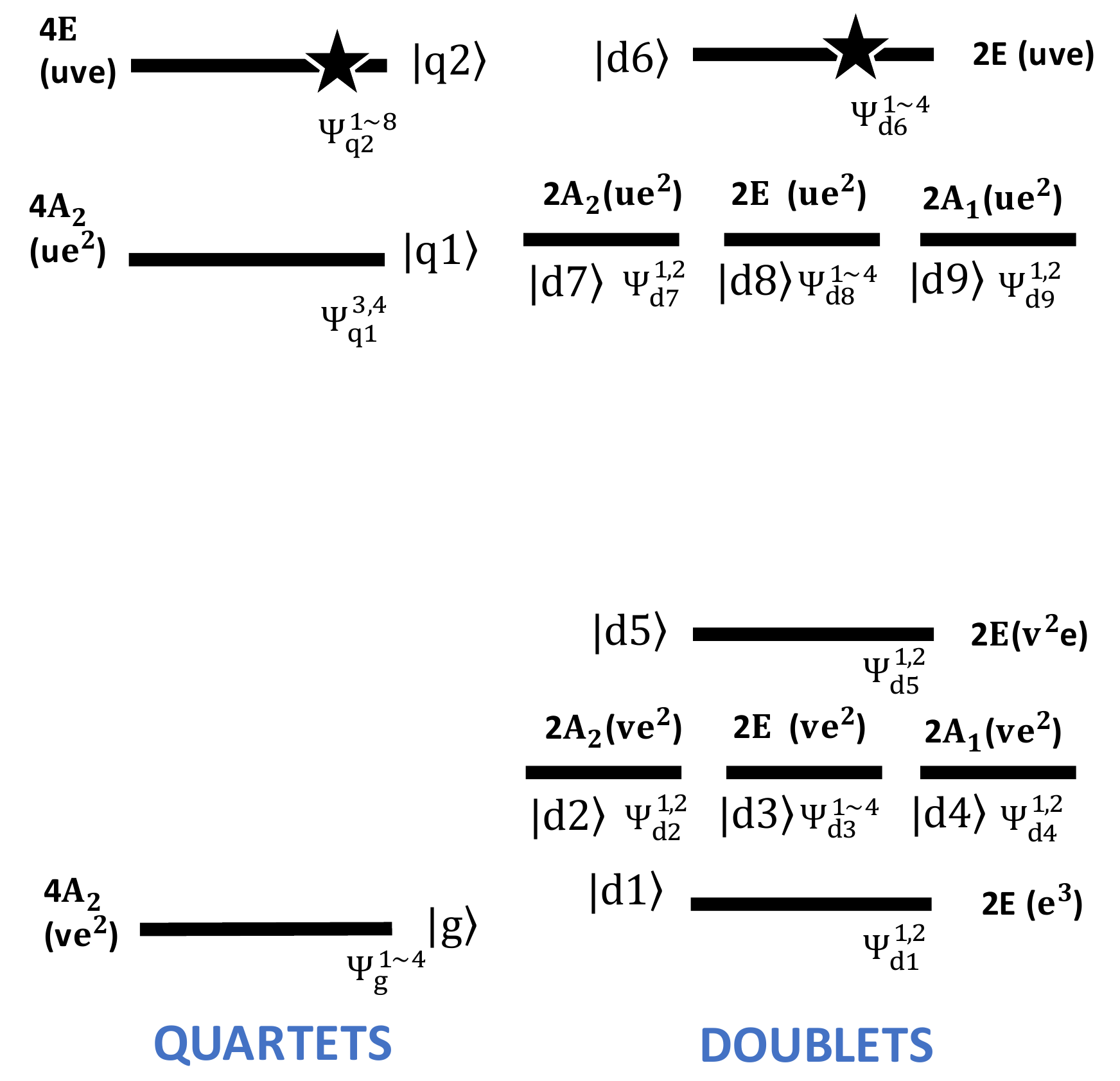} 
\caption{Electronic configuration characterized  by two different total spin numbers. The level spacing is meant to convey qualitatively our current understanding of the ordering of the states. The $d7,d8,d9$ doublets can be mapped from $d2,d3,d4$ under $v\rightarrow u$ orbital transformation (similar  to mapping $g$ to $q1$), they are plotted horizontally together for brevity.  The $\star$ symbols, which are only on the second excited quartet (q2) and the sixth doublet (d6),  indicate the natural mixture of wave functions  incurred by spin-orbit coupling.  }
\label{energy_level}
\end{figure} 

There are two inequivalent vacancy sites in SiC, one  hexagonal (h) and one quasi-cubic (k) for the $\text{V}_{\text{Si}}$ \cite{JanzenPhyB2009}. The local symmetry of  $\text{V}_{\text{Si}}$ in both cases is described by the ${C}_{{3v}}$ point group \cite{DresselhaussBook} (see Appendix A for more details). 
Based on the ${C}_{{3v}}$ projection formula, we can find the symmetry adapted many body  wave functions (i.e., three body in the holes picture) in terms of the single-particle symmetry adapted molecular orbitals, i.e., $e_x$, $e_y$, $v$ and $u$. 
This was done in Ref. \cite{SoykalPRB2016} to find most, but not all, of the states. Here we find the additional states, four doublets labelled $d6-d9$, which are crucial for the intersystem crossing of the defect. All states are presented in Appendix A and shown in Fig.\ref{energy_level}.

The spin-orbit coupling (SOC), which  couples the symmetry-adapted wave functions both within the degenerate manifolds and from different manifolds,  is expressed as :
\begin{equation}
\label{eq:SOC}
H_{SOC} = \sum_j  \vec{l_j} \cdot \vec{s_j} 
\end{equation}
where the $\vec{l}$ and $\vec{s}$ are orbital/spin angular momentum operators and the summation index $j$ is on different particles. We found the SOC mixes the wave functions within $q2$ and $d6$ only, and the mixed wave functions (all labelled by prime hereafter) are: $\{\Psi'^{(1-8)}_{q2} \} = \{ ( \Psi^2_{q2}-i\Psi^1_{q1})/\sqrt{2}, (\Psi^4_{q2}+i\Psi^3_{q1})/\sqrt{2} , \Psi^7_{q1}, \Psi^8_{q1}, \Psi^5_{q1}, \Psi^6_{q1}, (\Psi^4_{q2}-i\Psi^3_{q1})/\sqrt{2},(\Psi^2_{q2}+i\Psi^1_{q1})/\sqrt{2} \},$ which were also derived in  previous work \cite{SoykalPRB2016} and  $\{ \Psi'^{(1-4)}_{d6} \}=\{ (-\Psi^1_{d6}+\Psi^2_{d6})/\sqrt{2}; (\Psi^1_{d6}+\Psi^2_{d6})/\sqrt{2};  (-\Psi^3_{d6}+\Psi^4_{d6})/\sqrt{2}; (\Psi^3_{d6}+\Psi^4_{d6})/\sqrt{2}\} $, which were not found before.   In the following context, we always use the mixed states and neglect the prime and star notation on them.

 \begin{widetext}
 
\squeezetable
\begin{table}
\caption{SOC between quartets and doublets (we used the SOC mixed  $q2$ and $d6$,  labeled as prime).}
\label{table:SOC}
\begin{ruledtabular}
\scalebox{0.90}{
\begin{tabularx}{\textwidth}{|c|cccc|cccc|cccccccc|}
 & $\Psi^1_{g} $ & $\Psi^2_{g} $ &$\Psi^3_{g} $  &$\Psi^4_{g} $ &$\Psi^1_{q1} $ & $\Psi^2_{q1} $ &$\Psi^3_{q1} $  &$\Psi^4_{q1} $  & $\Psi^{\prime1}_{q2}$ &$\Psi^{\prime2}_{q2}$&$\Psi^{\prime3}_{q2}$&$\Psi^{\prime4}_{q2}$&$\Psi^{\prime5}_{q2}$&$\Psi^{\prime6}_{q2}$&$\Psi^{\prime7}_{q2}$&$\Psi^{\prime8}_{q2}$ \\
\hline
 $  \Psi^1_{d1}  $ &$-\lambda _{\text{$\bot$1}} $ & $ -\lambda _{\text{$\bot$1}} $ & $ 0 $ & $ 0 $ & $ -\lambda _{\text{$\bot$2}} $ & $ -\lambda _{\text{$\bot$2}} $ & $ 0 $ & $ 0 $ & $ 0 $ & $ 0 $ & $ 0 $ & $ 0 $ & $ 0 $ & $ 0 $ & $ 0 $ & $ 0$ \\
 $  \Psi^2_{d1}  $ &$i \lambda _{\text{$\bot$1}} $ & $ -i \lambda _{\text{$\bot$1}} $ & $ 0 $ & $ 0 $ & $ i \lambda _{\text{$\bot$2}} $ & $ -i \lambda _{\text{$\bot$2}} $ & $ 0 $ & $ 0 $ & $ 0 $ & $ 0 $ & $ 0 $ & $ 0 $ & $ 0 $ & $ 0 $ & $ 0 $ & $ 0$ \\
 $  \Psi^3_{d1}  $ &$0 $ & $ 0 $ & $ \frac{i \lambda _{\text{$\bot$1}}}{\sqrt{3}} $ & $ -\frac{\lambda _{\text{$\bot$1}}}{\sqrt{3}} $ & $ 0 $ & $ 0 $ & $ \frac{i \lambda _{\text{$\bot$2}}}{\sqrt{3}} $ & $ -\frac{\lambda _{\text{$\bot$2}}}{\sqrt{3}} $ & $ 0 $ & $ 0 $ & $ 0 $ & $ 0 $ & $ 0 $ & $ 0 $ & $ 0 $ & $ 0$ \\
$  \Psi^4_{d1}  $ &$ 0 $ & $ 0 $ & $ -\frac{i \lambda _{\text{$\bot$1}}}{\sqrt{3}} $ & $ -\frac{\lambda _{\text{$\bot$1}}}{\sqrt{3}} $ & $ 0 $ & $ 0 $ & $ -\frac{i \lambda _{\text{$\bot$2}}}{\sqrt{3}} $ & $ -\frac{\lambda _{\text{$\bot$2}}}{\sqrt{3}} $ & $ 0 $ & $ 0 $ & $ 0 $ & $ 0 $ & $ 0 $ & $ 0 $ & $ 0 $ & $ 0$ \\
$  \Psi^1_{d2(d7)}   $ &$ 0 $ & $ 0 $ & $ 0 $ & $ 0 $ & $ 0 $ & $ 0 $ & $ 0 $ & $ 0 $ & $ 0 $ & $ 0 $ & $ \pm \frac{i \lambda _{\text{$\bot$2}}}{3 \sqrt{2}} $ & $  \mp \frac{i \lambda _{\text{$\bot$2}}}{3 \sqrt{2}} $ & $ 0 $ & $ 0 $ & $ 0 $ & $   \frac{i \lambda _{\text{$\bot$2}}}{\mp\sqrt{3}}$ \\
$  \Psi^2_{d2(d7)}   $ &$ 0 $ & $ 0 $ & $ 0 $ & $ 0 $ & $ 0 $ & $ 0 $ & $ 0 $ & $ 0 $ & $ 0 $ & $ 0 $ & $\pm  \frac{\lambda _{\text{$\bot$2}}}{3 \sqrt{2}} $ & $\pm  \frac{\lambda _{\text{$\bot$2}}}{3 \sqrt{2}} $ & $ 0 $ & $ 0 $ & $  \frac{i \lambda _{\text{$\bot$2}}}{\mp\sqrt{3}} $ & $ 0 $\\
$ \Psi^1_{d3(d8)}  $ &$ 0 $ & $ 0 $ & $ 0 $ & $ 0 $ & $ 0 $ & $ 0 $ & $ 0 $ & $ 0 $ & $ \pm i \lambda _{\text{$\bot$2}} $ & $ \pm \lambda _{\text{$\bot$2}} $ & $ 0 $ & $ 0 $ & $ 0 $ & $ 0 $ & $ 0 $ & $ 0 $\\
$  \Psi^2_{d3(d8)}  $ &$ 0 $ & $ 0 $ & $ 0 $ & $ 0 $ & $ 0 $ & $ 0 $ & $ 0 $ & $ 0 $ & $ \mp i \lambda _{\text{$\bot$2}} $ & $ \mp \lambda _{\text{$\bot$2}} $ & $ 0 $ & $ 0 $ & $ 0 $ & $ 0 $ & $ 0 $ & $ 0$ \\
$ \Psi^3_{d3(d8)}  $ &$ 0 $ & $ 0 $ & $ 0 $ & $ 0 $ & $ 0 $ & $ 0 $ & $ 0 $ & $ 0 $ & $ 0 $ & $ 0 $ & $ 0 $ & $ 0 $ & $ 0 $ & $  \frac{\sqrt{2}\lambda _{\text{$\bot$2}}}{\pm\sqrt{3}} $ & $ 0 $ & $ 0 $\\
$ \Psi^4_{d3(d8)} $ &$ 0 $ & $ 0 $ & $ 0 $ & $ 0 $ & $ 0 $ & $ 0 $ & $ 0 $ & $ 0 $ & $ 0 $ & $ 0 $ & $ 0 $ & $ 0 $ & $ \frac{\sqrt{2}\lambda _{\text{$\bot$2}}}{\mp\sqrt{3}}$ & $ 0 $ & $ 0 $ & $ 0 $\\
$ \Psi^1_{d4(d9)}  $ &$ 0 $ & $ 0 $ & $ {\frac{4i}{\sqrt{6}}} \text{$\lambda_\parallel$}(0) $ & $ 0 $ & $ 0 $ & $ 0 $ & $ 0({\frac{4i}{\sqrt{6}}} \text{$\lambda_\parallel$}) $ & $ 0 $ & $ 0 $ & $ 0 $ & $\pm \frac{\lambda _{\text{$\bot$2}}}{\sqrt{6}} $ & $ \mp \frac{\lambda _{\text{$\bot$2}}}{\sqrt{6}} $ & $ 0 $ & $ 0 $ & $ 0 $ & $\pm \lambda _{\text{$\bot$2}} $\\
$  \Psi^2_{d4(d9)}   $ &$ 0 $ & $ 0 $ & $ 0 $ & $ {\frac{4i}{-\sqrt{6}}}\text{$\lambda _\parallel$}(0) $ & $ 0 $ & $ 0 $ & $ 0 $ & $ 0({\frac{-4i}{\sqrt{6}}} \text{$\lambda_\parallel$}) $ & $ 0 $ & $ 0 $ & $ \pm \frac{i \lambda _{\text{$\bot$2}}}{\sqrt{6}} $ & $ \pm \frac{i \lambda _{\text{$\bot$2}}}{\sqrt{6}} $ & $ 0 $ & $ 0 $ & $ \mp \lambda _{\text{$\bot$2}} $ & $ 0 $\\
$  \Psi^1_{d5}  $ & $-i \lambda^{\dagger}_{\text{$\bot$1}} $ & $ -i \lambda^{\dagger}_{\text{$\bot$1}} $ & $ 0 $ & $ 0 $ & $ 0 $ & $ 0 $ & $ 0 $ & $ 0 $ & $ 0 $ & $ 0 $ & $ 0 $ & $ 0 $ & $ 0 $ & $ 0 $ & $ 0 $ & $ 0 $\\
$  \Psi^2_{d5}  $ & $\lambda^{\dagger}_{\text{$\bot$1}} $ & $ -\lambda^{\dagger}_{\text{$\bot$1}} $ & $ 0 $ & $ 0 $ & $ 0 $ & $ 0 $ & $ 0 $ & $ 0 $ & $ 0 $ & $ 0 $ & $ 0 $ & $ 0 $ & $ 0 $ & $ 0 $ & $ 0 $ & $ 0 $\\
$  \Psi^3_{d5}  $ &$ 0 $ & $ 0 $ & $ \frac{\lambda^{\dagger}_{\text{$\bot$1}}}{\sqrt{3}} $ & $ \frac{i \lambda^{\dagger}_{\text{$\bot$1}}}{\sqrt{3}} $ & $ 0 $ & $ 0 $ & $ 0 $ & $ 0 $ & $ 0 $ & $ 0 $ & $ 0 $ & $ 0 $ & $ 0 $ & $ 0 $ & $ 0 $ & $ 0 $\\
$  \Psi^4_{d5}  $ &$ 0 $ & $ 0 $ & $ -\frac{\lambda^{\dagger}_{\text{$\bot$1}}}{\sqrt{3}} $ & $ \frac{i \lambda^{\dagger}_{\text{$\bot$1}}}{\sqrt{3}} $ & $ 0 $ & $ 0 $ & $ 0 $ & $ 0 $ & $ 0 $ & $ 0 $ & $ 0 $ & $ 0 $ & $ 0 $ & $ 0 $ & $ 0 $ & $ 0 $\\
$   \Psi'^1_{d6}  $ &$ \frac{\sqrt{3} }{2} \lambda^{\dagger}_{\text{$\bot$2}} $ & $ \frac{\sqrt{3}}{2}  \lambda^{\dagger}_{\text{$\bot$2}} $ & $ 0 $ & $ 0 $ & $ \frac{ \sqrt{3} }{2}\lambda^{\dagger}_{\text{$\bot$1}} $ & $ \frac{\sqrt{3}}{2}  \lambda^{\dagger}_{\text{$\bot$1}} $ & $ 0 $ & $ 0 $ & $ 0 $ & $ 0 $ & $ 0 $ & $ 0 $ & $ 0 $ & $ 0 $ & $ 0 $ & $ 0 $\\
$   \Psi'^2_{d6}  $ &$ \frac{\lambda^{\dagger}_{\text{$\bot$2}}}{-2\sqrt{3}} $ & $ -\frac{\lambda^{\dagger}_{\text{$\bot$2}}}{2\sqrt{3}} $ & $ 0 $ & $ 0 $ & $ \frac{\lambda^{\dagger}_{\text{$\bot$1}}}{2\sqrt{3}} $ & $ \frac{\lambda^{\dagger}_{\text{$\bot$1}}}{2\sqrt{3}} $ & $ 0 $ & $ 0 $ & $ 0 $ & $ 0 $ & $ 0 $ & $ 0 $ & $ \frac{2 \text{$\lambda_\parallel$}}{3} $ & $ 0 $ & $ 0 $ & $ 0 $ \\
$   \Psi'^3_{d6} $ &$ \frac{i \sqrt{3}}{-2}  \lambda^{\dagger}_{\text{$\bot$2}} $ & $ \frac{i \sqrt{3} }{2} \lambda^{\dagger}_{\text{$\bot$2}} $ & $ 0 $ & $ 0 $ & $ \frac{i \sqrt{3}}{-2}  \lambda^{\dagger}_{\text{$\bot$1}} $ & $ \frac{i \sqrt{3}}{2}  \lambda^{\dagger}_{\text{$\bot$1}} $ & $ 0 $ & $ 0 $ & $ 0 $ & $ 0 $ & $ 0 $ & $ 0 $ & $ 0 $ & $ 0 $ & $ 0 $ & $ 0  $\\
$   \Psi'^4_{d6}  $ &$ \frac{i\lambda^{\dagger}_{\text{$\bot$2}} }{2\sqrt{3}}  $ & $ \frac{-i \lambda^{\dagger}_{\text{$\bot$2}}}{2\sqrt{3}} $ & $ 0 $ & $ 0 $ & $ \frac{-i \lambda^{\dagger}_{\text{$\bot$1}}}{2\sqrt{3}} $ & $ \frac{i\lambda^{\dagger}_{\text{$\bot$1}}}{2\sqrt{3}}   $ & $ 0 $ & $ 0 $ & $ 0 $ & $ 0 $ & $ 0 $ & $ 0 $ & $ 0 $ & $ \frac{2 \text{$\lambda_\parallel$}}{3} $ & $ 0 $ & $ 0 $\\
\end{tabularx}
}
\end{ruledtabular}
\end{table}

\end{widetext}

\section{Intersystem Crossing }

Intersystem crossing (ISC) is a non-radiative mechanism of  transition between  electronic states with different spin numbers.  For the   $\text{V}_{\text{Si}}$ in SiC, the total spin is either  $S=\frac{3}{2}$ (spin quartets)  or  $S=\frac{1}{2}$  (spin doublets) as shown in Fig. \ref{energy_level}. Optical pumping alone cannot realize ISC, as it does not couple  states with different total spin or spin projection. The  strongest  spin changing mechanism  is SOC (spin-spin interactions are weaker and will be neglected in our calculation).  The SOC not only mixes wave functions within the sub manifold, but also, importantly, couples wave functions from quartets and doublets. To represent the coupling strength, by using the Wigner-Eckart theorem to reduce the result, we can simplify the SOC between any two wave functions  to three parameters  $\lambda_{\parallel}=-i\bra{E}|O^{A_2} |\ket{E}$,  $\lambda_{\perp,1}=\frac{-i}{\sqrt{2}}\bra{A_1(v)}|O^{E}|\ket{E}$ and $\lambda_{\perp,2}=\frac{-i}{\sqrt{2}}\bra{A_1(u)}|O^{E}|\ket{E}$ (where $O^J$ is an operator belonging to the $J$ representation of $C_{3v}$) only, which are quantified in \cite{DohertyPhyRep2013}.  The symmetry of orbital and spin angular momentum operators are: $(l_x , l_y, s_x, s_y)  \mapsto O^E, (l_z,s_z)    \mapsto O^{A_2}$.
The SOC between quartets  and doublets are in Table \ref{table:SOC}.  One should note that in Table \ref{table:SOC} we use the mixed wave functions for $q2$ and $d6$ and they have the prime symbols. The actual transition dynamics also contain the phonon-assisted transition (we use the term `phonon' somewhat loosely in this work to refer to both delocalized and localized vibrational modes). Therefore, in this section we focus on how phonons couple to electronic transitions in the ISC process. 
We follow a similar approach to  Goldman et al. \cite{GoldmanPRB2015,GoldmanPRL2015}, while we note that the ISC mechanism in $\text{V}_{\text{Si}}$ is much more complex than in the NV center due to the the larger total spin number and the higher number of energy levels, which enable a larger number of transitions. 

The SOC and  phonon coupling can be combined to describe the ISC transition rate, therefore  each electronic state in the transitional process should be generally dressed by the vibrational state, which we use to label the total state. For example, $\ket{q1, \nu_0}$ represents the first excited quartet in its ground vibrational  state. For the ISC starting from a specific quartet to a target doublet, the direct ISC rate is: 
\begin{equation}
\Gamma^{(1)} \propto  |\lambda_{\bot(1,2)}|^2\sum_n |\braket{\chi_0}{\chi'_{\nu_n}}|^2\delta({\nu_n-\Delta}), \label{eq:Gamma1}
\end{equation}
where, $\propto$ represents equivalence up to numerical factors from SOC among specific quartet and target doublets, which can be found in Table \ref{table:SOC}.  States $\ket{\chi_0}$ and $\ket{\chi'_{\nu_n}}$  are the ground vibrational state of the quartet and an excited vibrational state of the target doublet respectively;  $\nu_n$  is the energy separating the excited vibrational level of the doublet and its ground vibrational state;  $\Delta$ is the energy difference between $q1$ and the target doublet when both are at their ground vibrational states ($\Delta=\epsilon_{q1}-\epsilon_{d}$). The above formula only captures the unexcited (ground) vibrational mode for  $q1$ while  an excited version can be derived similarly (Eq. \eqref{eq:gamma-1-general}). Generally, the strength of the ISC depends on the energy difference $\Delta$ between initial and final states; the ISC will be weak if $\Delta$ is too large for the vibrational modes to overcome. In terms of the energy separation to the excited quartets,  we can classify the doublets into two groups $\{ d6,d7,d8,d9 \}$ and $\{ d1,d2,d3,d4\}$ depending on their orbital configurations. 

\begin{figure} 
\centering
\includegraphics[scale=0.42]{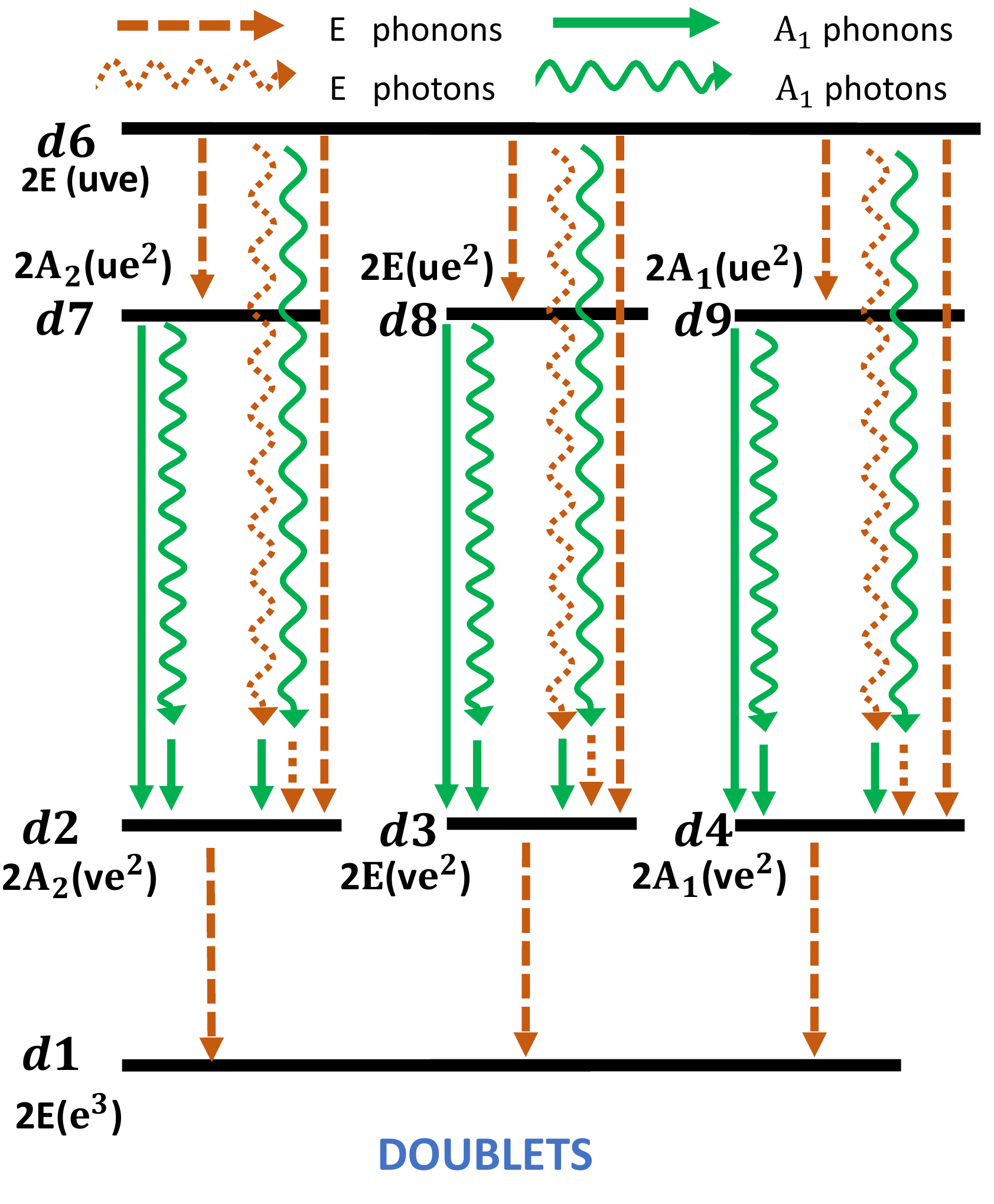}
\caption{Selection rules for the inter-doublet relaxation process, which is accompanied by the emission of a phonon or a photon (or both). Photon emission process is represented by curly lines and phonon process by straight lines for  $A_1$ (green/solid) and  $E$ (brown/dashed or dotted). For transitions with large energy difference, phonon  process alone is unlikely. The more physical case involves a combination of photon and phonon process. } 
\label{phonons among doublets }
\end{figure}

Generally, phonons do couple different electronic states. 
 We can represent the   electron-phonon interaction as:
\begin{equation}
H_{\text{e-ph}}=\sum_{p,k}V_{\text{ph}}^p\delta_{p,k}(a^{\dagger}_{p,k}+a_{p,k}), \label{eq:Hamiltonian}
\end{equation}
where the projectors on single orbitals (Appendix B) give rise to the  projector $V_{\text{ph}}^p$ among symmetry-adapted wave functions, and  $\delta_{p,k}$ is the phonon coupling rate  (also shown  in  Eq. \eqref{eq:Hstrain}); $a_{p,k}$ and $a_{p,k}^{\dagger}$ are the  annihilation and creation operators with wave vector $k$ and polarization $p$. 
In Fig. \ref{phonons among doublets }, based on the application of selection rules, we show the permitted phononic transitions among some representative doublets in terms of phonon symmetry type. The possible phononic  transitions within doublets assist the dynamics of ISC, e.g. in Section IV,  two doublets $d6$ and $d4$ contribute to the ISC dynamics  to realize spin polarization. Phonons of $E$ symmetry couple $d6$ and $d4$, and within the interaction Hamiltonian we find the  projectors for the symmetry-adapted wave functions to be:

\begin{eqnarray}
V_{\text{ph}}^1 &=& -i\frac{\sqrt{3}}{4} \ket{\Psi'^1_{d6}}\bra{\Psi^1_{d4}}+i\frac{\sqrt{3}}{4} \ket{\Psi'^2_{d6}}\bra{\Psi^1_{d4}} \nonumber \\
&+&i\frac{\sqrt{3}}{4} \ket{\Psi'^3_{d6}}\bra{\Psi^2_{d4}} - i\frac{\sqrt{3}}{4} \ket{\Psi'^4_{d6}}\bra{\Psi^2_{d4}} \label{eq:Vph1}\\
V_{\text{ph}}^2 &=  & \frac{\sqrt{3}}{4} \ket{\Psi'^1_{d6}}\bra{\Psi^1_{d4}}-\frac{\sqrt{3}}{4} \ket{\Psi'^2_{d6}}\bra{\Psi^1_{d4}} \nonumber\\
&+& \frac{\sqrt{3}}{4} \ket{\Psi'^4_{d6}}\bra{\Psi^2_{d4}} -\frac{\sqrt{3}}{4} \ket{\Psi'^5_{d6}}\bra{\Psi^2_{d4}} \label{eq:Vph2}.  
\end{eqnarray}

Once the phononic density of states is calculated, the above projectors along with Eq. (\ref{eq:Gamma1}) can quantify the rate. 
ISC through other doublets not accessible by SOC can occur through an indirect  (2nd order) process.  For instance, $q1$ and $d4$ are not directly coupled by SOC, but they are indirectly coupled  as $q1$ $\rightarrow$ $d6$ $\rightarrow$ $d4$. The $q1$ $\rightarrow$ $d6$ transition is enabled by SOC.

The second part of the transition can occur through relaxation via emission of either phonons, photons, or both. The case of only phonon-mediated relaxation, schematically shown in Fig. \ref{ISC channel in Vsi}(a), $E$ phonons are involved:
\begin{equation*}
\sum_m\ket{q1,\chi_m}\xrightarrow[\text{}]{\text{SOC}} \sum_n \ket{d6,\chi_n}\xrightarrow[\text{}]{\text{phonon}} \sum_l \sum_{p,q}\sum_{\pm1}\ket{d4,\chi_l}.
\end{equation*} 
Using the second order Fermi golden rule, in this scenario we obtain the second order ISC rate as (see Appendix B):
\begin{widetext}
\begin{eqnarray}
\Gamma^{(2)} & \propto & |\lambda_{\bot2}|^2 \sum_{m,l,p,q} \left[ \left | \sum_{n}  \frac{\tilde{\delta}_{pk} \braket {\chi_n} {\chi_m}  \sqrt{n_{p,q}+1} \braket {\chi_l}{\chi^+_n}  }{\Delta_6+\nu_m-\nu_n-\omega_{p,q}} \right |^2\delta( \Delta_4+\nu_m-\nu_n-\omega_{p,q} ) \right . \nonumber   \\
& + & \left . \left | \sum_n \frac{\tilde{\delta}_{pk} \braket{\chi_{n}}{\chi_{m}}  \sqrt{n_{p,q}}\braket{\chi_l}{\chi^-_n} }{\Delta_6+\nu_m-\nu_n+\omega_{p,q}} \right |^2   \delta(\Delta_4+\nu_m-\nu_n+\omega_{p,q}) \right],
\label{eq:Gamma2}
\end{eqnarray}
\end{widetext}
where $\Delta_{(4,6)} = \epsilon_{q1}-\epsilon_{d(4,6)}$.

The relaxation between doublets can also include a spontaneous photon emission, with either $A_1$ or $E$ symmetry (polarization along $z$ or in the $xy$ plane respectively), as indicated in Fig. \ref{ISC channel in Vsi}(b). Such a process is most likely the dominant mechanism for relaxation between doublets from the group $\{ d6,d7,d8,d9 \}$ and those from $\{ d1,d2,d3,d4\}$, compared to a purely phonon-driven scenario, due to the large energy difference  between the groups. This is analogous to the intersystem crossing and spin polarization cycle in the NV center in diamond, where an optical transition between singlets has been observed \cite{RogersNJP2008,KehayiasPRB2013}. 
 
\section{Spin polarization via optically driven ISC}

The optically-assisted spin polarization dynamics have been analyzed in the NV center, and the associated microscopic mechanisms have been identified and quantified \cite{GoldmanPRB2015,GoldmanPRL2015,ThieringPRB2018}. Here, we use our model from the previous section to construct similar spin-polarization protocols for $\text{V}_{\text{Si}}$. As the quartets have two excited  manifolds, i.e., the first excited quartet $q1$ and the second excited quartet $q2$, ISC can occur either between $q1$ and doublets or between $q2$ and doublets. We first explore the first ISC from $q1$.

\subsection{First spin polarization channel: from $q1$ to $g$ }
\begin{figure} 
\centering
\includegraphics[scale=0.42]{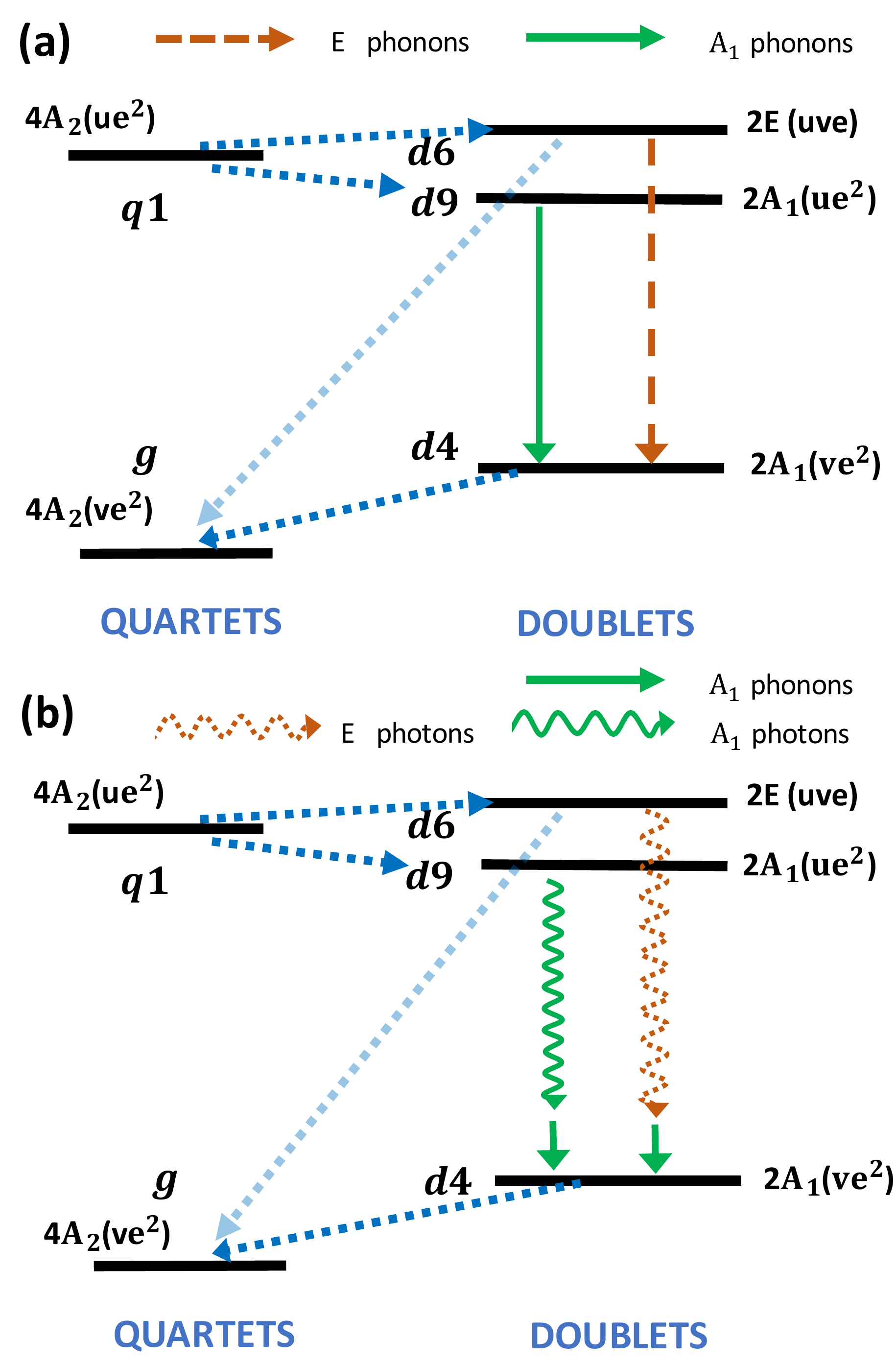}
\caption{ISC channel  starting from $q1$ involving different photon and phonon emissions. States $d6$ and $d9$ couple to $d4$ by (a) phonons or  (b) spontaneous photon emission along with phonon emission. States $d9$ and $d4$ are coupled with $A_1$ symmetry  and $d6$ and $d4$ are coupled  with $E$ symmetry. }
\label{ISC channel in Vsi}
\end{figure}

Based on the calculated spin-orbit coupling matrix elements from Table \ref{table:SOC}, we find that the  first ISC  from $q1$ occurs to doublets $d1$, $d6$ and $d9$, while other doublets are not directly  coupled to $q1$ (see Fig.\ref{ISC channel in Vsi}).

Following the method in Section III  the corresponding $q1$ to $d6$ transition rate is:  

\begin{equation}
\label{eq:ISCd6}
\Gamma_{q1\text{-}d6}  \propto  |\lambda_{\perp2}|^2\sum^{\in {\{d6\}}}_n |\braket{\chi_0}{\chi_{\nu_n}}|^2\delta({\nu_n-\Delta_{q1,d6}})
\end{equation}
where, $\braket{\chi_0}{\chi_{\nu_n}}$ is the overlap  of states between phonon ground states and excited states.  Similarly, the $d4$ to $g$ transition rate is:
\begin{equation}
\label{eq:ISCd4g}
\Gamma_{d4\text{-}g}  \propto   |\lambda_{\parallel}|^2\sum^{\in {\{d4\}}}_n |\braket{\chi_0}{\chi_{\nu_n^{\prime}}}|^2\delta({\nu_n^{\prime}-\Delta_{d4,g}}),
\end{equation}
This transition rate is nonzero only for the $|S_z|=\frac{1}{2}$ $g$ states.

The same approach can be applied to $d1$ to obtain a similar equation.  However the transition from $q1$ to $d6$ is presumably much stronger than that from $q1$ to $d1$ as both $d6$ and $q1$ states have $uve$ orbital configurations and, more importantly, it is  energetically much closer to $q1$, whereas the vibrational modes of $d1$ cannot compensate for the large $\Delta_{q1,d1}$, making the transition rate much weaker. Moreover, the $q1{\rightarrow} d1{\rightarrow} g$ and  $q1{\rightarrow} d6{\rightarrow} g$  ISC channels  feature a spin-conserving mechanism, i.e., the spin projection of $g$ states will be preserved after the cycle. Therefore there does not exist a single doublet that can be used in a three-level model to polarize the ground state. This conclusion is consistent with experimental results \cite{FuchsNatCom2015}. This phenomenon can be  explained by the similar symmetry of $g$ ($ve^2$) and $q1$ ($ue^2$) states:  both $v$ and $u$ have  $A_1$ symmetry and the $g$ can be mapped to $q1$ by changing orbital $v$ to $u$, so for a specific doublet, the selection rule applies equivalently for ground and $q1$ wave functions.  In Ref.~\cite{FuchsNatCom2015}, a four-level model was proposed to explain the transition. Here, based on our work, we can assign either $d3$ or $d9$ to their metastable level and the population from the metastable levels can be removed either optically or through phonon or photon assisted decay to lower doublets.

For a complete, microscopic model of spin polarization through the excited manifold $q_1$, we consider all the possible transitions between the high energy doublets and those with lower energy. Among the high energy doublets, $d9$ ($d6$) can couple to  $d4$  by $A_1$ ($E$) symmetry relaxation, as discussed above and illustrated in Fig. \ref{ISC channel in Vsi}. We consider different possible combinations of photon and  phonon symmetry for a transition with a given symmetry. For example, for a transition with $E$ character, one possibility is that $E$(total)=$A_1$(photon) $\otimes$ $E$(phonon) and another is $E$(total)=$E$(photon) $\otimes$ $A_1$(phonon). We believe that the former option is more likely, as it resembles the NV case. In fact, we speculate that even a similar vibrational mode as in NV-diamond may be involved in the case of V$_\text{Si}$; from the experimental results of the W\"urzburg group, who found that the optimal excitation energy to maximize photoluminescence from the defect is 172 meV above the ZPL \cite{HainJAP2014}, and comparing to a vibrational mode found in NV-diamond of 169 meV that plays a key role in the relaxation between singlets \cite{KehayiasPRB2013}, we assign the $A_1$ phonon accompanying the photon emission to this mode. Note that because this mode has been found to be very localized in NV-diamond and to mainly involve the basal carbons (and not the nitrogen), it is quite likely that essentially the same mode exists in V$_\text{Si}$. As in diamond, this mode is outside the phonon spectrum of the bulk SiC material \cite{ProtikMatTodPhys2017}. In fact, in the data of Fuchs et al. \cite{FuchsNatCom2015} there is evidence for additional localized vibronic modes at lower frequencies (although one has to be careful in interpreting the data, as these are ensemble experiments and could involve signal from other defects); such (quasi)localized lower-frequency modes are consistent with the bulk phonon spectrum of SiC \cite{ProtikMatTodPhys2017}, which has a bandgap ($\sim$70-90 meV), a feature that is distinct from diamond.

There are two low-lying doublet states, $d1$ and $d4$, that directly connect to the ground state manifold. Since we do not know the ordering of these states, we will consider two models, each corresponding to one of these doublets directly relaxing to the ground state. 

We begin by analyzing the case of direct relaxation of $d4$ to $g$. As $d4$ only couples to the $|S_z|=\frac{1}{2}$ in the $g$ quartet (Eq. \eqref{eq:ISCd4g}),  by using  $d6$ and $d4$ as the intermediate states,  we find a way that the $q1$ state with  spin $|S_z|=\frac{3}{2}$  can transition to the $g$ states with $|S_z|=\frac{1}{2}$  while the reverse transition does not occur, realizing  a spin-flipping process:
\begin{equation}
\label{eq:gamma1-flip}
\gamma_{|\frac{3}{2}|\rightarrow|\frac{1}{2}|} \gg \gamma_{|\frac{1}{2}|\rightarrow|\frac{3}{2}|}  \approx 0 .
\end{equation}

\begin{figure} 
\centering
\includegraphics[scale=0.41]{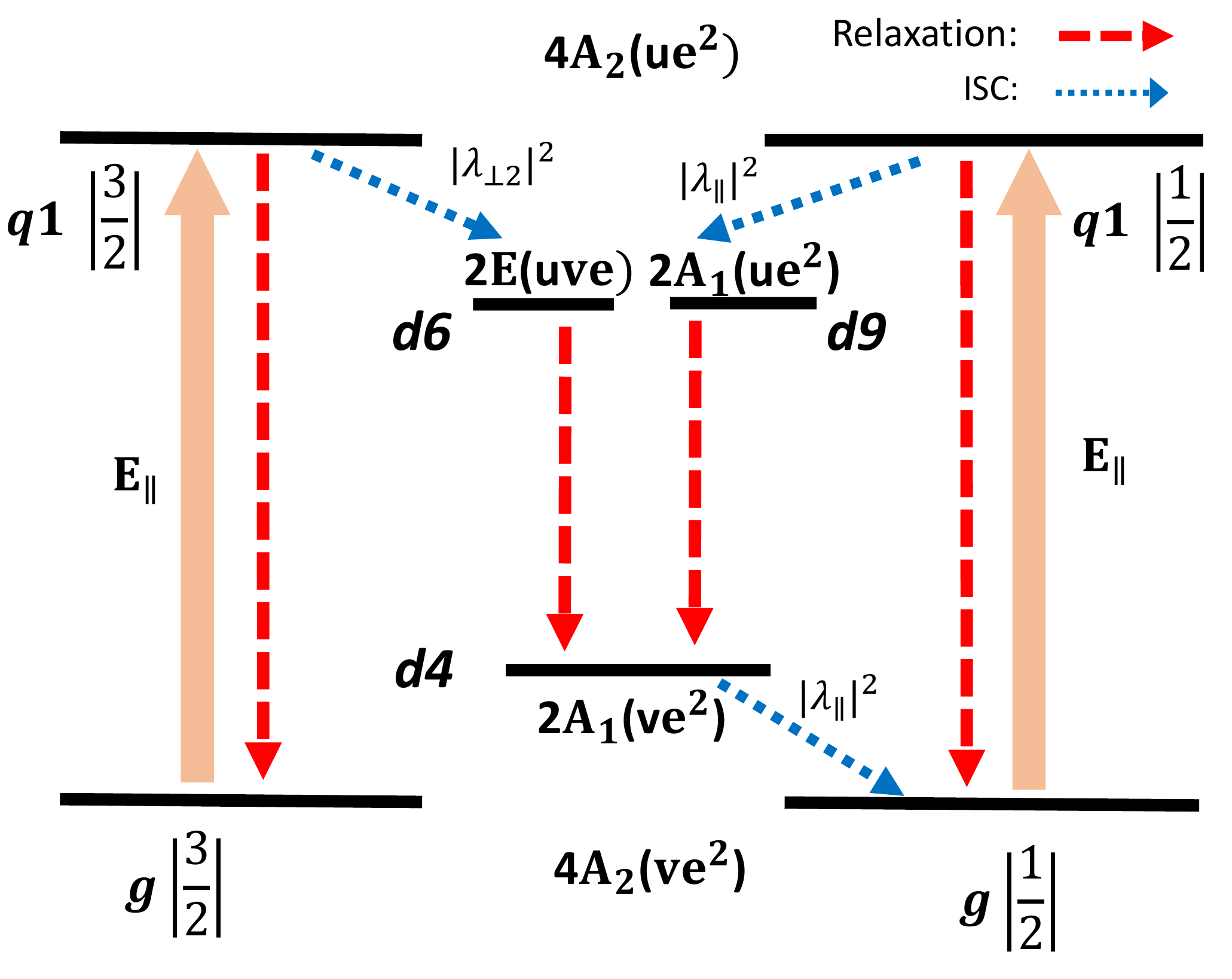}
\caption{ Spin polarization protocol for $g-q1$ quartets by optical pumping. The $E_{\parallel}$ type optical pumping drives the ground to the excited quartet ($q1$).  ISC couples $q1$ and $d6$, spontaneous photon emission takes $d6$ to $d4$, which is also coupled to the ground quartet.  Due to the strong spontaneous emission between the two quartets  and the large  $q1-d1$  energy separation for ISC , the indirect transition via  $d1$ can be neglected. }
\label{spin_polar_q1}
\end{figure}

Based on the spin-flipping ISC from $q1$ to $d4$,  $d6$ and $d9$ doublets, we construct the first spin polarization protocol.  The doublets involved could be effectively reduced to $d4$, $d6$ and $d9$ (Fig. \ref{spin_polar_q1}). 

\begin{figure} 
\centering
\includegraphics[scale=0.3]{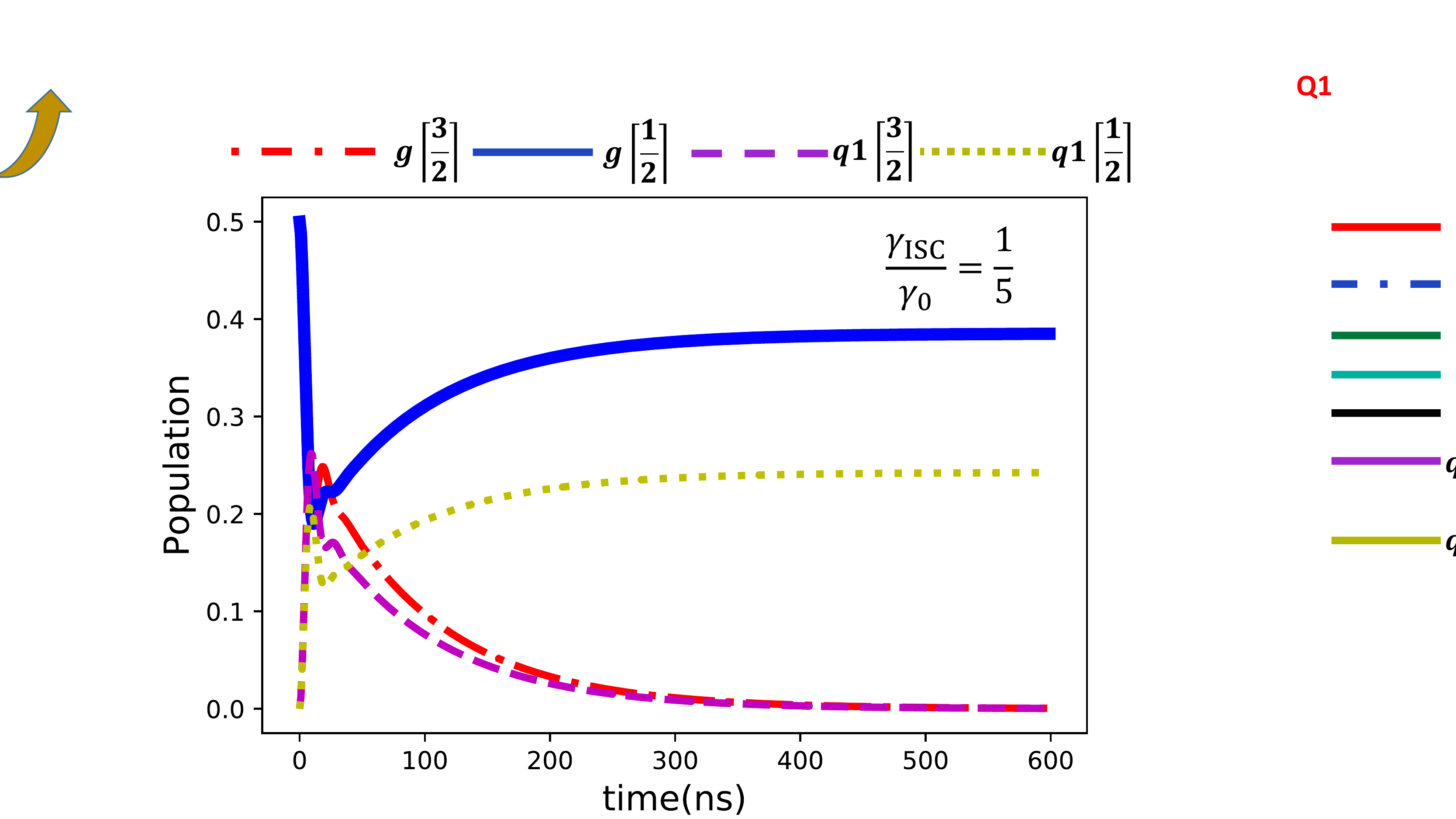}
\caption{ Spin polarization dynamics for the first protocol by using  optical pumping between $g$ and $q1$ quartets and assuming that the decay from doublet $d4$ dominates relaxation back into the ground state. The ratio of the ISC and spontaneous emission rates is taken to be $\frac{1}{5}$. Quartet $g$ with $|S_z|=\frac{1}{2}$ (blue/solid line) will by populated  asymptotically. Once the laser is off, it is close to 100\% populated. }
\label{lindblad_q1}
\end{figure}

The states evolve  according to the Lindblad equation:
\begin{equation}
\label{eq:lindblad}
\dot{\rho}(t)= -i[H,\rho(t)]+\sum_k (L_k\rho(t)L^\dagger_k-\frac{1}{2}\{L^\dagger_kL_k,\rho(t) \}), 
\end{equation}
where the model includes two states ($|S_z|=\frac{3}{2} $ and $\frac{1}{2}$) from each quartet $g$ and $q1$ and one state from each of the doublets $d4$, $d6$, and $d9$, hence it is  seven dimensional. We consider resonant drive between  $g$ and $q1$, and define $\Omega$ to be the Rabi frequency. The Lindblad operators $L_k$, which are given in Appendix C, contain the ISC rates and spontaneous emission rate.  We fix the optical drive strength $\Omega=$1/$6.1 $ ns$^{-1}$ and the spontaneous emission rate $\gamma_0 \approx \Omega$. Using an ISC rate value comparable to what was deduced in Ref. \cite{FuchsNatCom2015}, we find that spin polarization can occur in several hundreds of nanoseconds, as shown in Fig. \ref{lindblad_q1}. (the steady state shows around 40\% population on the excited $|S_z|=\frac{1}{2}$, which, once the pumping is turned off, is transferred to  ground $|S_z|=\frac{1}{2}$ under spin conserving spontaneous emission). Then the final polarization of $|S_z|=\frac{1}{2}$ within the ground quartet should approach 100$\%$. The timescale of several hundreds of ns is consistent with experiment \cite{WidmannNatureMat2014,NiethammerPRAP2016}.

\begin{figure} 
\centering
\includegraphics[scale=0.42]{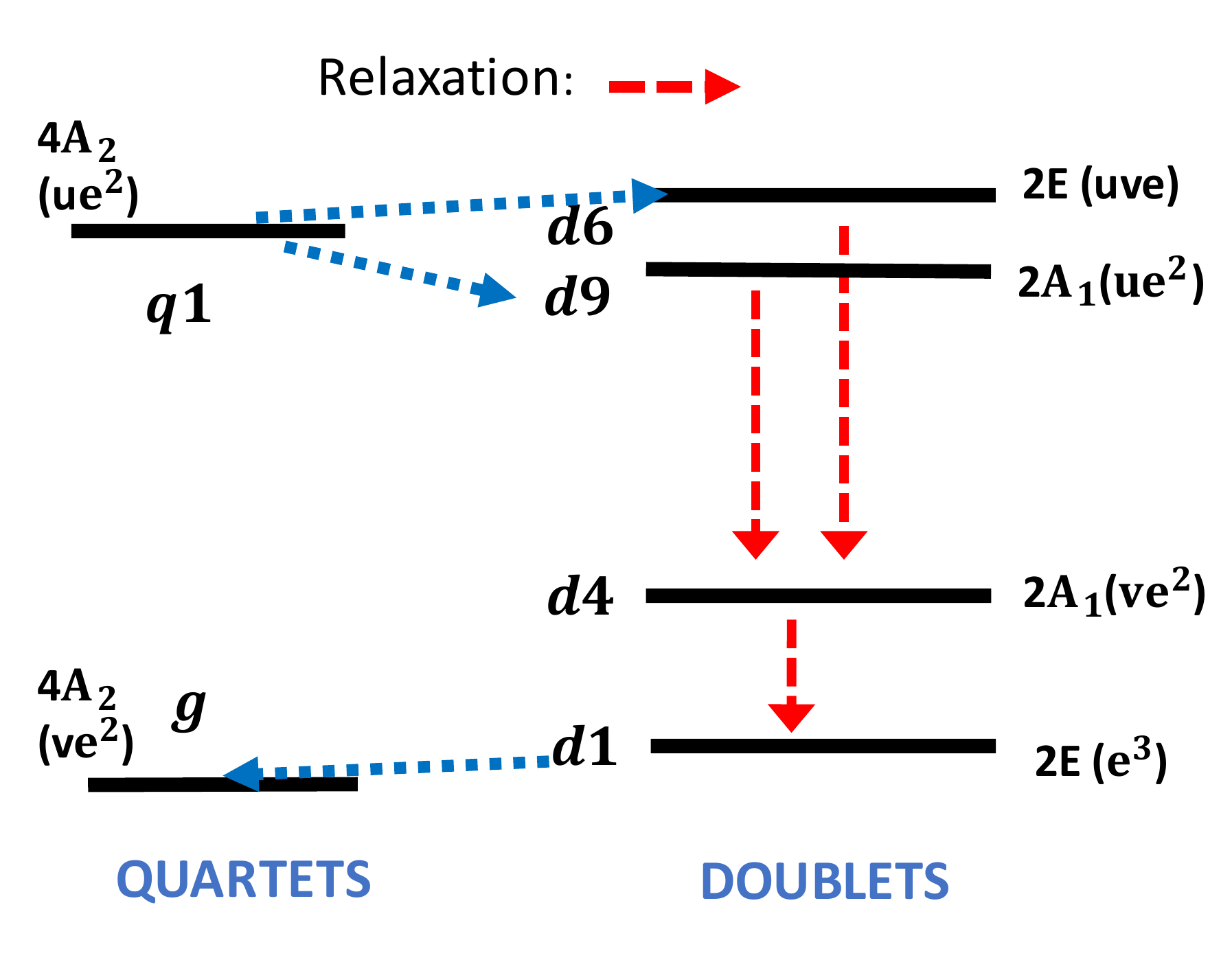}
\caption{ISC channel  starting from $q1$ involving different photon and phonon emissions. States $d6$ and $d9$ couple to $d4$ by (a) phonons or  (b) spontaneous photon emission along with phonon emission. States $d9$ and $d4$ are coupled with $A_1$ symmetry  and $d6$ and $d4$ are coupled  with $E$ symmetry. }
\label{ISC channel in Vsi-d1}
\end{figure}

\begin{figure} 
\centering
\includegraphics[scale=0.32]{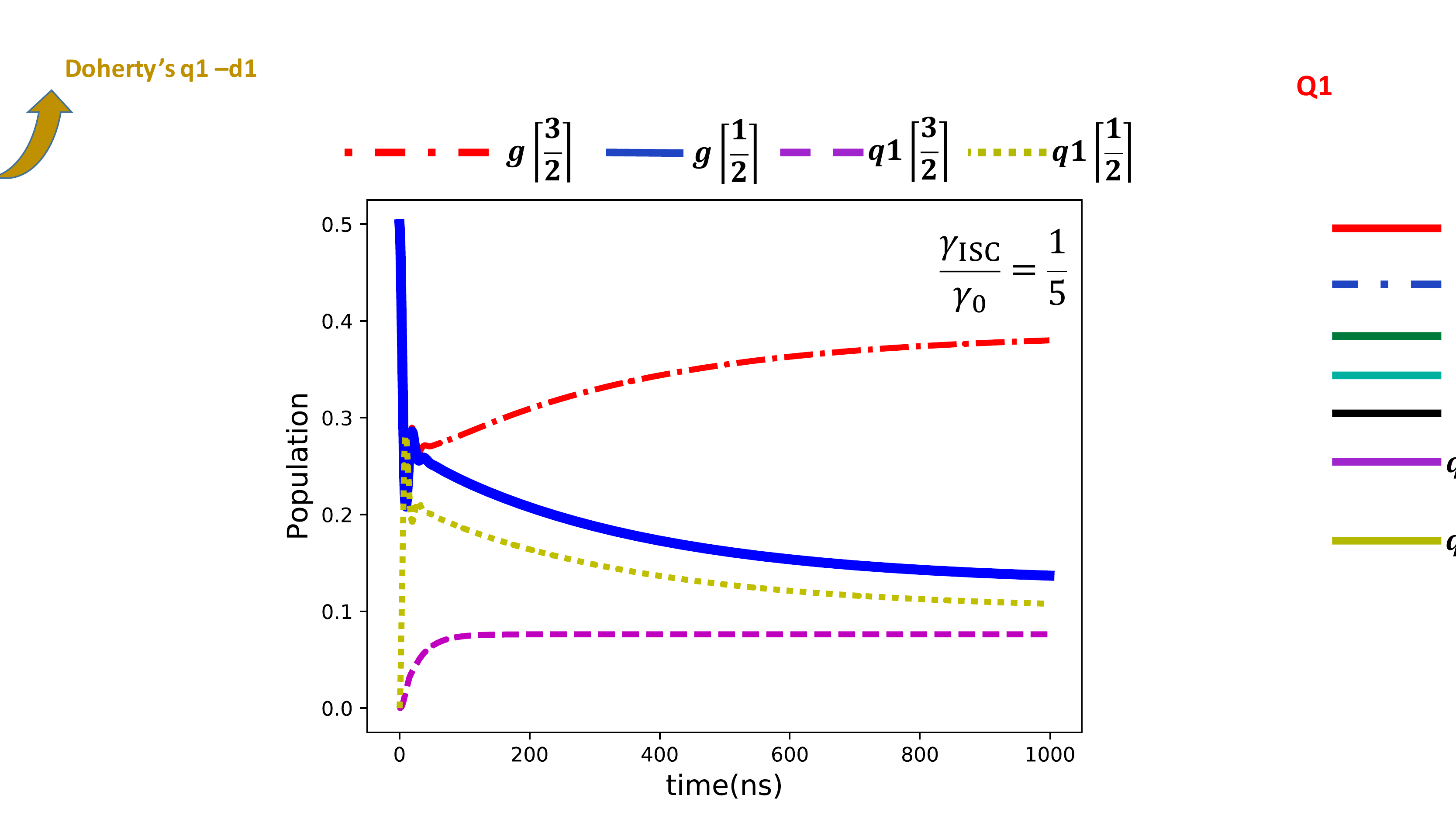}
\caption{ Spin polarization dynamics for the first protocol  by using  optical pumping between   $g$ and $q1$ quartets  and assuming that the decay from doublet $d1$ dominates relaxation back into the ground state. The ratio of the ISC and spontaneous emission rates is taken to be $\frac{1}{5}$. Quartet $g$ with $|S_z|=\frac{3}{2}$ (red/dotted-dashed line) will by predominantly populated. }
\label{lindblad_q1_d1}
\end{figure}

An alternative scenario to what is described above is that $d4$ first relaxes to $d1$, which in turn relaxes to the ground state. This mechanism assumes that $d4$ has higher energy, something that is not known yet. Because of the limited information about these doublets, we consider this channel as a possibility as well, as shown in Fig. \ref{ISC channel in Vsi-d1}. Solving a Lindblad equation as before, in this case, we find that the other spin projection states ($|S_z|$=3/2) are polarized, albeit not fully, since a considerable fraction of the population remains in the $|S_z|$=1/2 states, see Fig. \ref{lindblad_q1_d1}. 

\subsection{Second spin polarization channel: from $q2$ to $g$}
\begin{figure}
\centering
\includegraphics[scale=0.50]{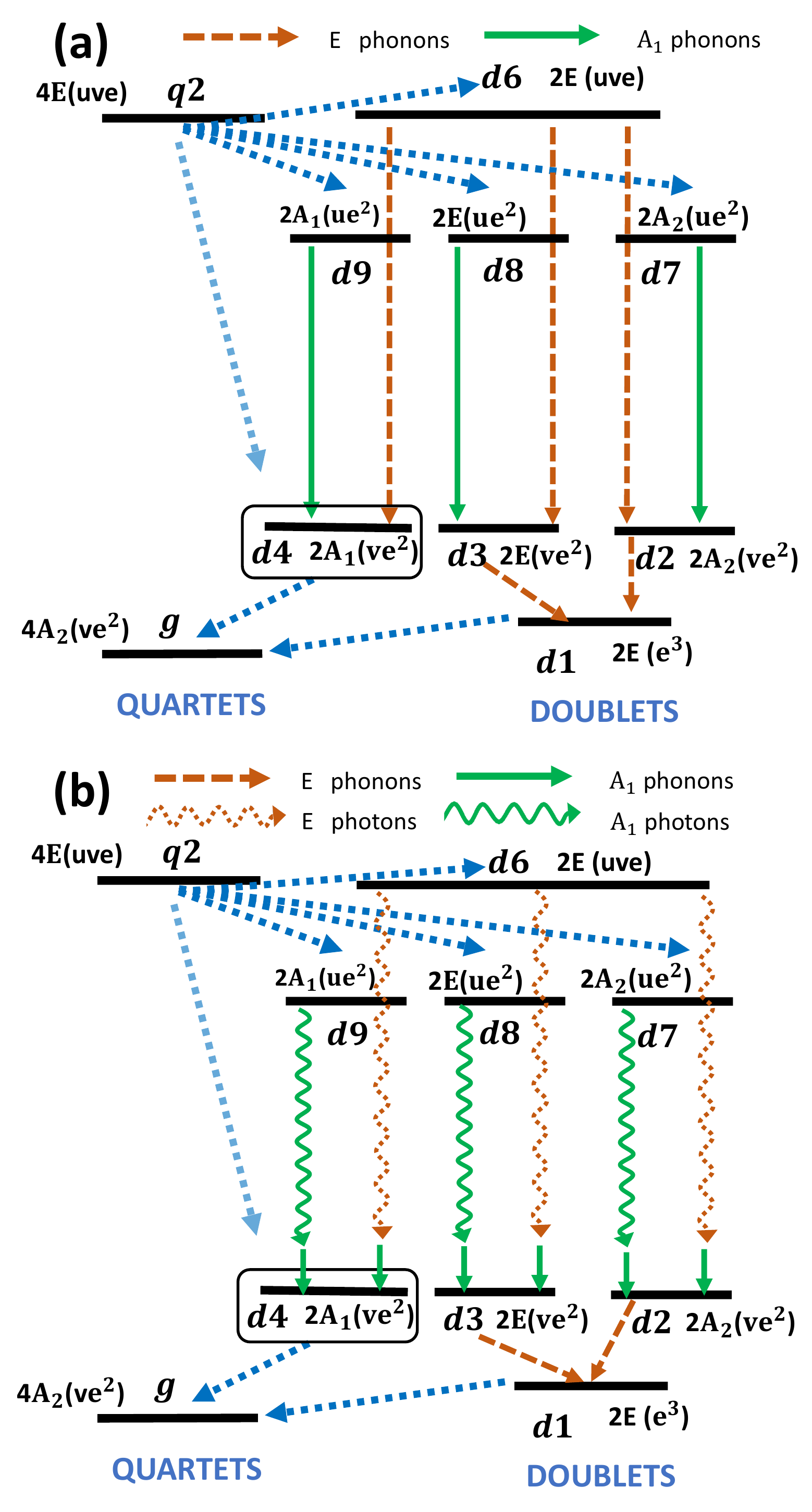}
\caption{Doublet $d4$ is the only state which couples to $q2$ and $g$ simultaneously and has spin flipping transitions, allowing for a simple three-state model of spin polarization. Starting from $q2$, a more likely channel involves intermediate states $d6$, $d7$, $d8$ and $d9$ and through phonons and optical spontaneous emission, these states can couple to $d2$, $d3$ and $d4$ respectively. Doublet $d6$ can couple to $d2$, $d3$ and $d4$. Both $d2$ and $d3$ relax to the $g$ quartet indirectly through $d1$. As in the $q1$ channel, we indicate (a) phonon-only processes and (b) photon-phonon combined processes, with the latter more likely to happen.} 
\label{fig:q2channel}
\end{figure}

ISC also occurs via the second excited quartet $q2$, and can also lead to ground-state spin polarization. The physics of the ISC from $q2$ is more complicated compared to that from $q1$. One qualitative difference between the two cases is that there exists a doublet ($d4$) which couples to $q2$ and $g$ simultaneously and has spin-flipping transitions. Therefore, we could construct a three-level model accordingly (Fig. \ref{fig:q2channel}). However, the energy conservation would require phonons that match the large frequencies of the transitions. Therefore, this model is less likely compared to a four- (or more) level model for spin polarization via $q2$. We find that all doublets in $\{ d6,d7,d8,d9 \}$ can couple to $q2$ directly and, due to their orbital configuration, we should not ignore any of them. As discussed above, $d7$, $d8$ and $d9$ can couple to  their isomorphic states $d2$, $d3$ and $d4$ respectively, by $A_1$ symmetry relaxation. As states $d2$, $d3$ and $d4$ share the same orbital configurations and therefore their energy difference should be comparatively small, $d6$ can couple to each of them  through $E$ relaxation. Again, we assume an $E$ photon and $A_1$ phonon as the more plausible combination, shown in Fig.\ref{fig:q2channel} (b). On the other hand, unlike $d4$, $d2$ and $d3$ do not couple to $g$ directly,  but indirectly through $d1$.  Therefore, the ISC and spin polarization protocol of $q2$ is quite complex, as is illustrated in Fig. \ref{fig:q2channel}. 
 
To explain the spin polarization mechanism, we need to specify how the spin-flipping process occurs among the complex ISCs. We demonstrate all the possible transitions in Fig. \ref{fig:q2channel} and compare their relative strengths. We can focus on the doublets that couple to $g$ quartets directly, i.e. $d1$ and  $d4$. We find that transitions from $\Psi'^{(3-6)}_{q2}$ to $g$ through $d4$ are spin conserving  and transitions from $\Psi'^{(7,8)}_{q2}$ to $g$ through $d4$  are spin flipping, which is in contrast to that in the first spin polarization protocol. The remaining ISCs within this protocol go through $d1$. We find that $d2$ and  $d3$ can couple to both $|S_z|=\frac{3}{2}$ and $|S_z|=\frac{1}{2}$ of $g$, hence  transitions via $d1$ are mixtures of spin conserving and spin flipping.  Next, we need  to compare the spin-flipping process with opposite directions:
\begin{eqnarray}
\label{eq:gamma2-flip}
\gamma_{|\frac{3}{2}|\rightarrow|\frac{1}{2}|} &=& \sum_i \gamma^i_{|\frac{3}{2}|\rightarrow|\frac{1}{2}|}   \nonumber \\
&=&\gamma^{_{(d2,d1)}}_{ _{|\frac{3}{2}|\rightarrow|\frac{1}{2}|}}+ \gamma^{_{(d3,d1)}}_{ _{|\frac{3}{2}|\rightarrow|\frac{1}{2}|}} +\gamma^{_{(d4)}}_{ _{|\frac{3}{2}|\rightarrow|\frac{1}{2}|}}  \\
\gamma_{|\frac{1}{2}|\rightarrow|\frac{3}{2}|} &=& \sum_i \gamma^i_{|\frac{1}{2}|\rightarrow|\frac{3}{2}|}   \nonumber \\
&=&\gamma^{_{(d2,d1)}}_{ _{|\frac{1}{2}|\rightarrow|\frac{3}{2}|}}+ \gamma^{_{(d3,d1)}}_{ _{|\frac{1}{2}|\rightarrow|\frac{3}{2}|}} 
\end{eqnarray}
where,  $\gamma^{_{(d2,d1)}}_{ _{|\frac{3}{2}|\rightarrow|\frac{1}{2}|}}$ for example, represents the transitions from  $|S_z|=\frac{3}{2}$ to $|S_z|=\frac{1}{2}$ going through $d2$ and $d1$. 
But comparing those two groups of  spin-flipping transitions is  challenging due to the complex paths they take and the difficulty of quantifying their strengths.  One  crucial example is the transition from $d6$ to $d4$ and that from $d9$ to $d4$:  even if we can express their transition  rates by referring  to equations in Section III, their relative ratio  requires the knowledge of the density of states of their vibrational modes. To the best of our knowledge, there are no first principles calculations available  from which to obtain these parameters. 

\begin{figure} 
\centering
\includegraphics[scale=0.60]{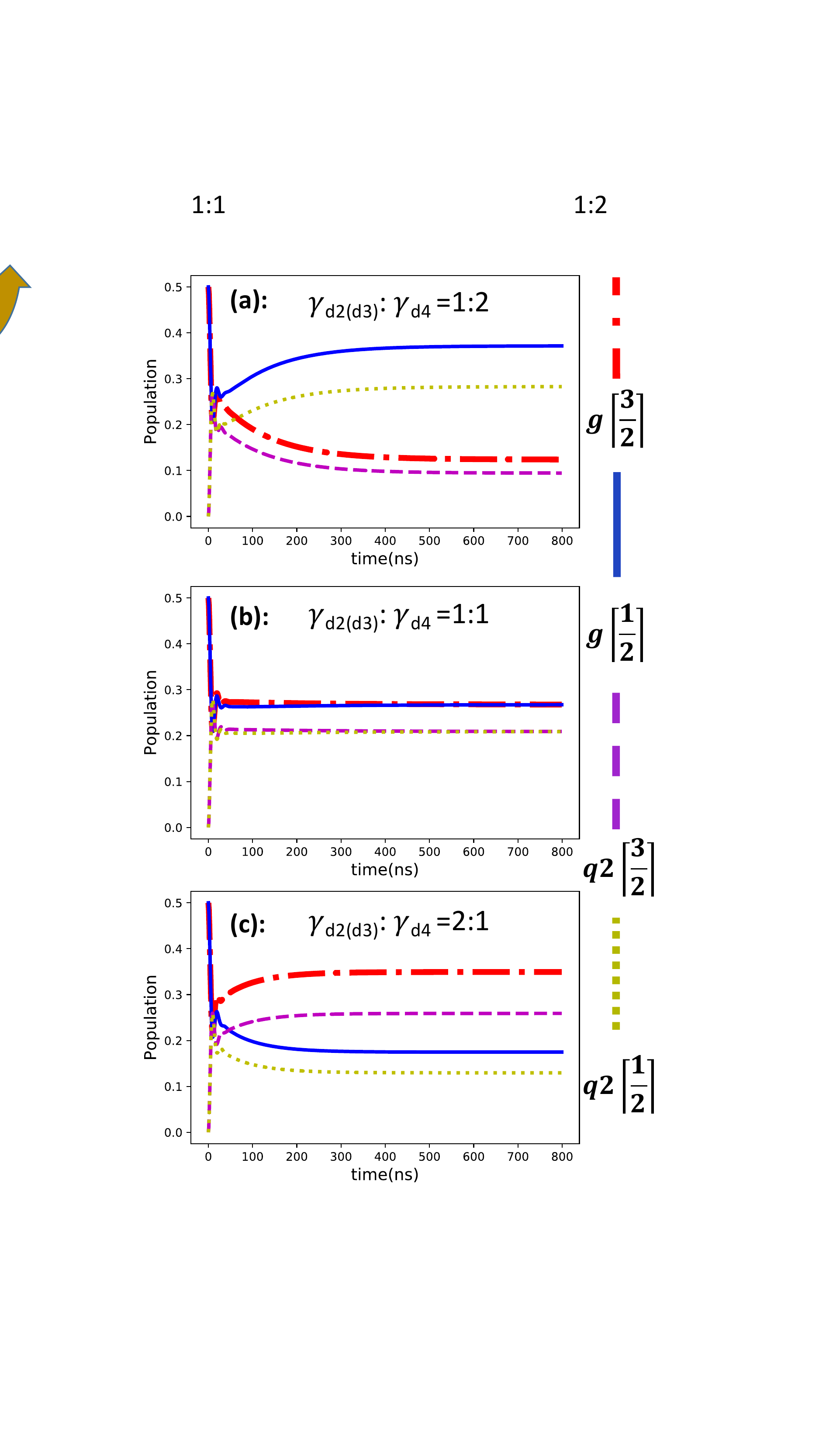}
\caption{Spin polarization dynamics when pumping $q2$. The ISC and spontaneous emission ratio is $\frac{1}{5}$. The ratio of the ISC rates to $d2(d3)$ and $d4$ is varied. (a) $\gamma_{d2(d3)}/\gamma_{d4}=$ 1:2, (b) $\gamma_{d2(d3)}/\gamma_{d4}=$ 1:1, and (c)  $\gamma_{d2(d3)}/\gamma_{d4}=$ 2:1. In cases (a) and (c), a different initial spin projection state is polarized, while case (b) represents the crossover point, where no spin polarization is obtained.} 
\label{lindblad_q2}
\end{figure}

In the absence of further inputs from \emph{ab initio} calculations, we simplify the model with some reasonable assumptions. We focus on the $d2$, $d3$ and $d4$ doublets and ignore the higher doublets as these three determine the coupling to the $g$ quartets. Following the same approach as the first spin polarization protocol, we use Lindblad equations to describe the dynamics of this model, where we vary the ISC rates to $d2$, $d3$ and $d4$. Interestingly, in this case the system can be polarized in either spin projection state, $|S_z|=\frac{1}{2}$ or $|S_z|=\frac{3}{2}$, depending on the relative strength of the rates, as shown in Fig \ref{lindblad_q2} (a) and (c) respectively. This can be due to the different SOC strengths between the $g$ quartets and the three doublets, where $d2$ and $d3$ preferentially relax to $|S_z|=\frac{3}{2}$, while $d4$ relaxes to $|S_z|=\frac{1}{2}$ only. When the rates exactly balance each other no polarization is generated, as shown in Fig \ref{lindblad_q2} (b). We note that $q2$ states  split under axial SOC \cite{SoykalPRB2016}, presumably with splittings in the GHz range \cite{EconomouNanoTech2016,BatalovPRL2009}, so in principle a spectrally narrow laser could realize selective pumping and create spin polarization irrespective of the relative rates. 

\begin{figure}
\centering
\includegraphics[scale=0.4]{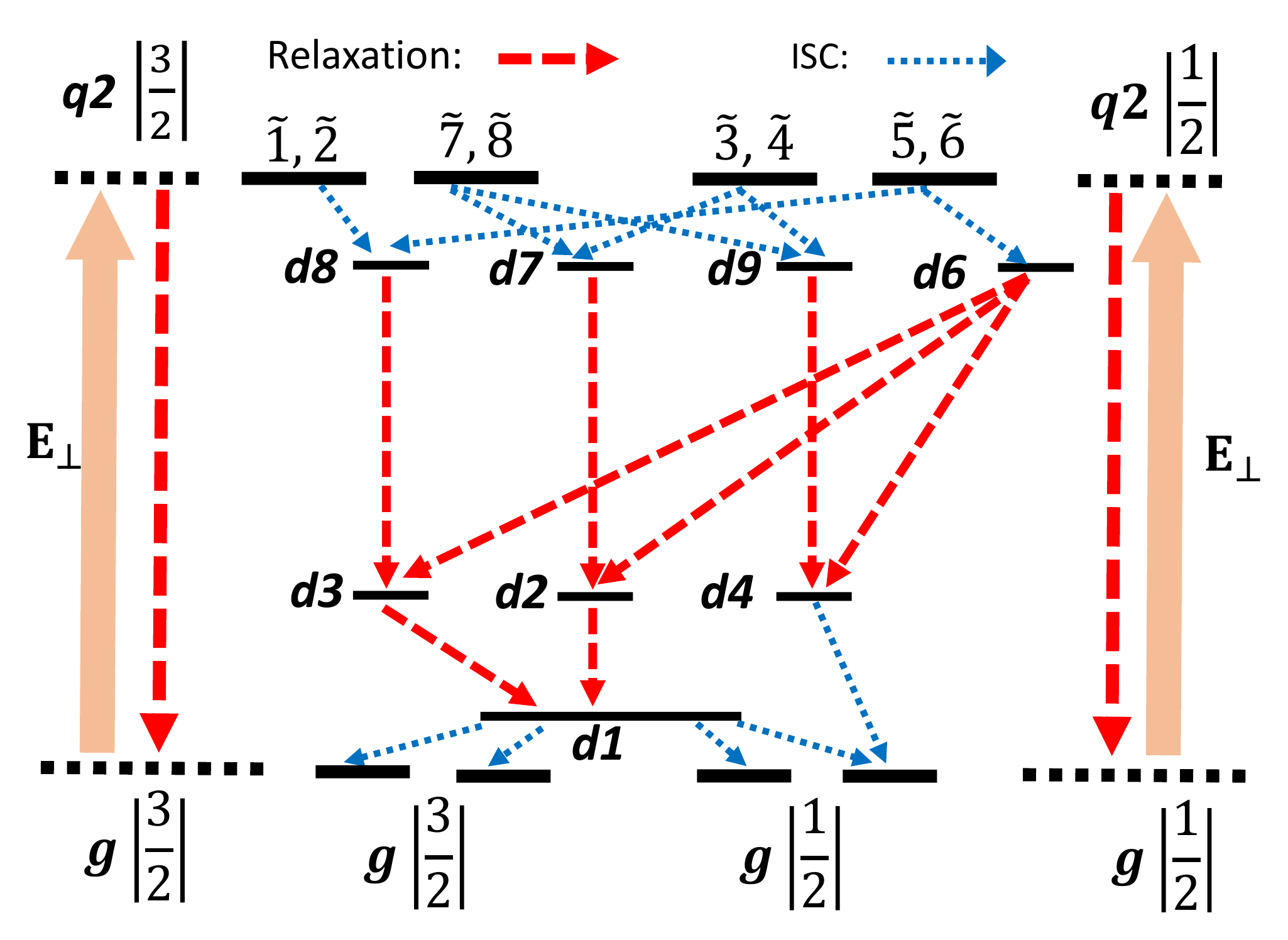}
\caption{ ISC from $q2$ to several doublets and finally to the ground quartet. For the doublets directly coupled to $g$, both $d4$  and $d1$ are  mixture of spin-flipping and spin conserving processes. The $E_{\bot} $ laser drives the system from $g$ to $q2$. }
\label{indirect_via_merge}
\end{figure} 

\section{Conclusion and Outlook}
In this paper, we studied the ISC dynamics by analyzing the SOC and the phonon coupling between symmetry-adapted many-particle states of $\text{V}_{\text{Si}}$ in SiC. We qualitatively analyzed the ISC among different spin manifolds and quantified the ratio of their rates. We analyzed two spin polarization protocols enabled by optical pumping, spin-orbit coupling, and interaction with phonons. The ISC mechanism through the second excited manifold ($q2$) is more complex as more doublets contribute to it. In general we find that both spin projections ($|S_z|=\frac{3}{2}$ or $|S_z|=\frac{1}{2}$) of the ground state manifold can be initialized, depending on the relative strength of inter-doublet relaxation rates and the relative ordering of the doublets. The two spin polarization channels discussed above can be distinguished by optical means.  According to selection rules, the ground quartet ($A_2$ symmetry) state  can be excited  to the first excited quartet ($A_2$ symmetry) by applying light polarized parallel to the c-axis $E_{\parallel}$, while the second excited quartet ($E$ symmetry)  by  light polarized perpendicular to the c-axis $E_{\perp}$. Our numerical simulations for the polarization process involve assumptions motivated by experimental results. Based on a comparison between experiments in NV centers in diamond \cite{KehayiasPRB2013} and in V$_{\text{Si}}$ defects in SiC \cite{HainJAP2014, FuchsNatCom2015} we speculate that a localized vibronic mode with frequency $\sim$170 meV is essentially the same mode and present in both defects. In the data of Fuchs et al. \cite{FuchsNatCom2015} there is evidence for additional localized vibronic modes at lower frequencies; such (quasi)localized lower-frequency modes are consistent with the bulk phonon spectrum of SiC \cite{ProtikMatTodPhys2017}, since they would lie in the bandgap (a feature that is not present in diamond). For a more quantitative theory and to lift some of the ambiguities, further input is needed from \emph{ab initio} calculations. In particular, calculations involving the vibrational modes and their coupling to the electronic defect levels would be particularly important. The ordering and spacing of the doublets, which requires calculations beyond DFT \cite{ChoiPRB2012} would also be an important input to further refine our model.

\begin{acknowledgments}
WD thanks Donovan Buterakos for useful discussions. SEE acknowledges support from the NSF, grant number DMR-1737921. MWD acknowledges support from the ARC, grant number DE170100169.
\end{acknowledgments}

\appendix

\section{Group Theory Information}
The basic ${C}_{{3v}}$ group (character table in Table \ref{table:character_table}), in conjunction with the SU(2) group for $\frac{1}{2}$ spin, forms the ${C}_{{3v}}$ double group \cite{DresselhaussBook} which gives the full description for the behavior of spinors under specific spatial symmetry.  The double group for spin $\frac{1}{2}$  is denoted as D$_\frac{1}{2}$, or $\Gamma_{E_{1/2}}$. A full  group symbol can be written as $\Gamma=\Gamma_o \otimes \Gamma_{E_{1/2}}$.

\subsubsection{Symmetry-adapted wave functions}
The ${V}_{{Si}}$   forms a local quantum few-body system with a discrete energy spectrum deep in the bandgap with four  single-particle molecular orbitals  -  $e_x$, $e_y$, $v$ and $u$. From those, the first two are degenerate and transform as $E$, while $v$ and $u$ transform as $A_1$.  The ${V}_{{Si}}$  has  five  electrons associated with it, four of which are from the four  carbon dangling bonds and one captured from environment. In this paper, we use the 3 holes picture  to find symmetry adapted many-body wave functions (filling 5 electrons in 8 states  $\{e_x,e_y,v,u  \} \otimes\{\uparrow,\downarrow \}$ is equivalent to filling  3 holes). The 3 holes can have a total spin of $\frac{3}{2}$ (quartet) or $\frac{1}{2}$ (doublet). The projector  can be scaled to the many particle situation. The modification is on the symmetry operation $P_R$. As the fermionic many-body wave functions are  conditioned  by Pauli exclusion principle and anti-symmetry of permutation, we need to construct a space transformation matrix $T$ - maps Hilbert space to antisymmetric space - and  transform the $P_R \rightarrow T P_R T^{\dagger}$. The symmetry-adapted total wave functions can be obtained by diagonalizing the projector and are listed (for brevity, single orbitals $e_x,e_y$  are represented by $x$, $y$) in Table \ref{table:wf_q} (16 quartets) and Table \ref{table:wf_d} (28 doublets). The decomposition of orbital and spinor symmetry type can be implemented by using the Clebsh-Gordan coefficients.
\begin{table}
\caption{Character table for $C_{3v}$ symmetry group}
\label{table:character_table}
\begin{ruledtabular}
\begin{tabular}{ |c |c c c c  c | } 
 ${C}_{{3v}}$  & E   & 2C$_3$  & 3$\sigma_{\nu}$ &   linear basis  & quadratic basis   \\
 \hline 
 A$_1$ & 1     &  1    &   1     &  z   &  $x^2+y^2, z^2$ \\ 
 A$_2$ & 1      & 1    &    -1   & $R_z$     &   \\ 
 E 	    & 2    &   -1  &    0    & (x,y)($R_x,R_y$) & ($x^2-y^2,2xy$)(xz,yz) \\ 
\end{tabular}
\end{ruledtabular}
\label{table:chartable}
\end{table}

\subsubsection{Projector and wave functions}
In group theory, the eigenvectors (denoted by $\Gamma_n j$)  relate the symmetry operator $P_R$ with its matrix representation denoted by $D^{\Gamma_n}(R)$ through the relation $P_R \ket{\Gamma_n \alpha} = \sum_j D^{\Gamma_n}(R)_{j\alpha} \ket{\Gamma_n j} $. With respect to the basis functions, the transformations can be described by the projection operators (or projectors) \cite{DresselhaussBook}   $P^{\Gamma_n}_{kl}$: $P^{\Gamma_n}_{kl} \ket{\Gamma_nl} \equiv  \ket{\Gamma_nk} $. The projector \cite{DresselhaussBook} is explicitly given in terms of the symmetry operators for the group by the relation: 
\begin{equation}
\label{eq:projector}
P^{\Gamma_n}_{kl}= \frac{l_n}{h} \sum_R D^{\Gamma_n}(R)^*_{kl} P_R, 
\end{equation}
where  $l_n$ and  $h$ are the dimension of $\Gamma_n$ and the rank of the group respectively. 

For our specific situation (to  fill three holes in $\{e_x,e_y,v,u  \} \otimes\{\uparrow,\downarrow \}$ orbitals), the symmetry operation is detailed as:
\begin{equation}
\label{eq:proj3holes}
 P_R(\text{3 holes})=\{ (\Gamma_E \otimes \Gamma_{1/2})\oplus(\Gamma_{A1} \otimes \Gamma_{1/2} )\oplus(\Gamma_{A1} \otimes \Gamma_{1/2} ) \}^{\otimes^3}
\end{equation}
Solving Eq. \eqref{eq:proj3holes} gives the exact wave functions, which are illustrated  in Table \ref{table:wf_q} and Table \ref{table:wf_d}, .

\begin{widetext}

\begin{table}
\caption{symmetry-adapted wave functions for spin quartets.}
\label{table:wf_q}
\begin{ruledtabular}
\begin{tabular}{ |c |  c| c| c|c|c| } 
 \hline
    Orbital   &  $m_S(\checkmark)$  & $\Gamma$   &  $\Gamma_o \otimes \Gamma_s $ &  symmetry-adapted wave functions(S=$\frac{3}{2}$)   & Label \\
    \hline
    \multirow{3}{2em} { $ve^2$ }          &  $\pm\frac{3}{2}$    &  $1E_{3/2}$  & $A_2 \otimes 2E_{3/2}$     &     $  \ket{|vxy+i\bar{v}\bar{x}\bar{y}}$        &  $\Psi_g^1$ \\ 
    & $\pm\frac{3}{2}$     &  $2E_{3/2}$  &   $A_2 \otimes 1E_{3/2}$     &    $  \ket{|vxy-i\bar{v}\bar{x}\bar{y}}$     &  $\Psi_g^2$ \\  
 ground  &    +$\frac{1}{2}$          &$E_{1/2}$  & $A_2 \otimes E_{1/2}$ &   $\ket{|vx\bar{y}+v\bar{x}y+\bar{v}xy}\sqrt{3}$   &   $\Psi_g^3$ \\ 
     & -$\frac{1}{2}$      &  $E_{1/2}$  & $A_2 \otimes E_{1/2}$ &  $\ket{|\bar{v}\bar{x}y+\bar{v}x\bar{y}+v\bar{x}\bar{y} }/\sqrt{3}$    & $\Psi_g^4$  \\
\hline
   \multirow{4}{2em} { $ue^2$ }          &  $\pm\frac{3}{2}$    &  $1E_{3/2}$  & $A_2 \otimes 2E_{3/2}$     &     $  \ket{|uxy+i\bar{u}\bar{x}\bar{y}}$        &  $\Psi_{q1}^1$ \\ 
    & $\pm\frac{3}{2}$     &  $2E_{3/2}$  &   $A_2 \otimes 1E_{3/2}$     &    $  \ket{|uxy-i\bar{u}\bar{x}\bar{y}}$     &  $\Psi_{q1}^2$ \\  
1st-excited   &    +$\frac{1}{2}$          &$E_{1/2}$  & $A_2 \otimes E_{1/2}$ &   $\ket{|ux\bar{y}+u\bar{x}y+\bar{u}xy}\sqrt{3}$   &   $\Psi_{q1}^3$ \\ 
     & -$\frac{1}{2}$      &  $E_{1/2}$  & $A_2 \otimes E_{1/2}$ &  $\ket{|\bar{u}\bar{x}y+\bar{u}x\bar{y}+u\bar{x}\bar{y} }/\sqrt{3}$    & $\Psi_{q1}^4$  \\
\hline
   \multirow{5}{2em} { $uve$ }          &  $+\frac{3}{2}$    &  $E_{1/2}$  & $E \otimes 1E_{3/2}$     &     $ \ket{|uvx}, \ket{|uvy}$        &  $\Psi_{q2}^1,\Psi_{q2}^2$ \\ 
  & $-\frac{3}{2}$     &  $E_{1/2}$  &   $E \otimes 2E_{3/2}$     &    $ \ket{|\bar{u}\bar{v}\bar{x}}, \ket{|\bar{u}\bar{v}\bar{y}}$     &  $\Psi_{q2}^3,\Psi_{q2}^4$ \\  
   &    \multirow{7}{2em} { $\pm\frac{1}{2}$ }            &$E_{1/2}$  &  \multirow{6}{4em} { $E\otimes E_{1/2}$ } &  $\ket{|(uv\bar{y}+u\bar{v}y+\bar{u}vy)+i(uv\bar{x}+u\bar{v}x+\bar{u}vx)}/\sqrt{6}$    &   $\Psi_{q2}^5$ \\ 
 2nd-excited      &        &  $E_{1/2}$  &   &  $\ket{|(\bar{u}\bar{v}y+\bar{u}v\bar{y}+u\bar{v}\bar{y})-i(\bar{u}\bar{v}x+\bar{u}v\bar{x}+u\bar{v}\bar{x})}/\sqrt{6}$    & $\Psi_{q2}^6$  \\
   &              &$1E_{3/2}$  &   &   $ \thead{  \{\ket{| (uv\bar{y}+u\bar{v}y+\bar{u}vy)-i(u\bar{v}\bar{y}+\bar{u}v\bar{y}+\bar{u}\bar{v}y ) }  \\ \ket{|-i(uv\bar{x}+u\bar{v}x+\bar{u}vx)+(u\bar{v}\bar{x}+\bar{u}v\bar{x}+\bar{u}\bar{v}x)} \}/2\sqrt{3}  }$   &   $\Psi_{q2}^7$ \\ 
     &        &  $2E_{3/2}$  &  &  $ \thead{  \{\ket{| (uv\bar{y}+u\bar{v}y+\bar{u}vy)+i(u\bar{v}\bar{y}+\bar{u}v\bar{y}+\bar{u}\bar{v}y ) }  \\ \ket{|-i(uv\bar{x}+u\bar{v}x+\bar{u}vx)-(u\bar{v}\bar{x}+\bar{u}v\bar{x}+\bar{u}\bar{v}x)} \}/2\sqrt{3}  }$    & $\Psi_{q2}^8$  \\     
 \hline
\end{tabular} 
\end{ruledtabular}
\end{table}

\begin{table}
\caption{symmetry-adapted wave functions for spin doublets.}
\label{table:wf_d}
\begin{ruledtabular}
\begin{tabular}{ |c |  c| c| c|c|c| } 
 \hline
    Orbital   &  $m_S(\checkmark)$  & $\Gamma$   &  $\Gamma_o \otimes \Gamma_s $ &  symmetry-adapted wave functions(S=$\frac{1}{2}$)   & Label \\
    \hline
    \multirow{4}{2em} { $e^3$ }          &  $+\frac{1}{2}$    &  $E_{1/2}$  &\multirow{4}{4em} { $E \otimes E_{1/2}$ }    &     $  \ket{| x\bar{x}y+iy\bar{y}x }/\sqrt{2}$        &  $\Psi_{d1}^1$ \\ 
    & $-\frac{1}{2}$     &  $E_{1/2}$  &         &    $  \ket{|\bar{x}x\bar{y}-i\bar{y}y\bar{x}}/\sqrt{2}$     &  $\Psi_{d1}^2$ \\  
   &    $\pm\frac{1}{2}$          &1$E_{3/2}$  &   &   $\ket{|({x}\bar{x}{y}-i{y}\bar{y}{x})-i(\bar{x}x\bar{y}-i\bar{y}y\bar{x})}/2$   &   $\Psi_{d1}^3$ \\ 
     & $\pm\frac{1}{2}$      &  $2E_{3/2}$  &   &  $\ket{|({x}\bar{x}{y}-i{y}\bar{y}{x})+i(\bar{x}x\bar{y}-i\bar{y}y\bar{x})}/2$     & $\Psi_{d1}^4$  \\
\hline
\hline
   \multirow{1}{2em} { $ve^2$ }          &  $+\frac{1}{2}$    &  $E_{1/2}$  & $A_2 \otimes E_{1/2}$     &     $  \ket{| vx\bar{y}+v\bar{x}y-2\bar{v}xy}/\sqrt{6}$        &  $\Psi_{d2}^1$ \\ 
    & $-\frac{1}{2}$     &  $E_{1/2}$  &   $A_2 \otimes E_{1/2}$     &    $  \ket{|\bar{v}\bar{x}y+\bar{v}x\bar{y}-2v\bar{x}\bar{y}}/\sqrt{6}$     &  $\Psi_{d2}^2$ \\  
    \hline
\multirow{4}{2em} { $ve^2$ }         & \multirow{3}{2em}{$\pm\frac{1}{2}$}     &  $1E_{3/2}$  &  \multirow{3}{4em} { $E \otimes E_{1/2}$}     &    $  \ket{|(vx\bar{y}-v\bar{x}y)-i(\bar{v}\bar{x}y-\bar{v}x\bar{y})+i(vx\bar{x}-vy\bar{y})-(\bar{v}\bar{x}x-\bar{v}\bar{y}y)     }/2\sqrt{2}$     &  $\Psi_{d3}^1$ \\  
    &      &  $2E_{3/2}$  &        &    $  \ket{|(vx\bar{y}-v\bar{x}y)+i(\bar{v}\bar{x}y-\bar{v}x\bar{y})+i(vx\bar{x}-vy\bar{y})+(\bar{v}\bar{x}x-\bar{v}\bar{y}y)     }/2\sqrt{2}$    &  $\Psi_{d3}^2$ \\  
    &      &  $E_{1/2}$  &         &    $  \ket{|(vx\bar{y}-v\bar{x}y)-i(vx\bar{x}-vy\bar{y})}/2$     &  $\Psi_{d3}^3$ \\  
    &      &  $E_{1/2}$  &        &    $  \ket{|(\bar{v}\bar{x}y-\bar{v}x\bar{y})+i(\bar{v}\bar{x}x-\bar{v}\bar{y}y) }/2$     &  $\Psi_{d2}^4$ \\  
    \hline
\multirow{1}{2em} { $ve^2$ }    &    +$\frac{1}{2}$          &$E_{1/2}$  & $A_1 \otimes E_{1/2}$ &   $\ket{|vx\bar{x}+vy\bar{y} }/\sqrt{2}$   &   $\Psi_{d4}^1$ \\ 
     & -$\frac{1}{2}$      &  $E_{1/2}$  & $A_1 \otimes E_{1/2}$ &  $\ket{| \bar{v}\bar{x}x+\bar{v}\bar{y}y }/\sqrt{2}$    & $\Psi_{d4}^2$  \\
\hline
\hline
   \multirow{5}{2em} { $v^2e$ }          &  $+\frac{1}{2}$    &  $E_{1/2}$  & \multirow{4}{4em}{$E \otimes E_{1/2}$ }    &     $ \ket{|v\bar{v}x-iv\bar{v}y}/\sqrt{2}$        &  $\Psi_{d5}^1$ \\ 
    & $-\frac{1}{2}$     &  $E_{1/2}$  &       &    $ \ket{|\bar{v}v\bar{x}+\bar{v}v\bar{y}}/\sqrt{2}$     &  $\Psi_{d5}^2$ \\  
   &     $\pm\frac{1}{2}$          &$1E_{3/2}$  &  &  $\ket{|(v\bar{v}x+iv\bar{v}y)+i(\bar{v}v\bar{x}-\bar{v}v\bar{y})}/2$    &   $\Psi_{d5}^3$ \\ 
     &   $\pm\frac{1}{2}$         &  $2E_{3/2}$  &   &  $\ket{|(v\bar{v}x+iv\bar{v}y)-i(\bar{v}v\bar{x}-\bar{v}v\bar{y})}/2$    & $\Psi_{d5}^4$  \\
  \hline
  \hline
   \multirow{5}{2em} { $uve$ }          &  $+\frac{1}{2}$    &  $E_{1/2}$  & \multirow{4}{4em}{$E \otimes E_{1/2}$ }    &     $ \ket{|i(uv\bar{x}+u\bar{v}x-2\bar{u}vx) +(uv\bar{y}+u\bar{v}y-2\bar{u}vy)}/2\sqrt{3}$        &  $\Psi_{d6}^1$ \\ 
    & $+\frac{1}{2}$     &  $E_{1/2}$  &       &    $ \ket{|i(\bar{u}vx+uv\bar{x}-2u\bar{v}x)+(\bar{u}vy+uv\bar{y}-2u\bar{v}y )}/2\sqrt{3}$     &  $\Psi_{d6}^2$ \\  
    & $-\frac{1}{2}$     &  $E_{1/2}$  &       &    $ \ket{|-i(\bar{u}v\bar{x}+\bar{u}\bar{v}x-2u\bar{v}\bar{x})+(\bar{u}v\bar{y}+\bar{u}\bar{v}y-2u\bar{v}\bar{y}) }/2\sqrt{3}$     &  $\Psi_{d6}^3$ \\  
   &     $-\frac{1}{2}$          & $E_{1/2}$  &  &  $ \ket{|-i( u\bar{v}\bar{x}+\bar{u}\bar{v}x-2\bar{u}v\bar{x})+(u\bar{v}\bar{y}+\bar{u}\bar{v}y-2\bar{u}v\bar{y}) }/2\sqrt{3}$    &   $\Psi_{d6}^4$ \\ 
  \hline
  \hline
   \multirow{1}{2em} { $ue^2$ }          &  $+\frac{1}{2}$    &  $E_{1/2}$  & $A_2 \otimes E_{1/2}$     &     $  \ket{| ux\bar{y}+u\bar{x}y-2\bar{u}xy}/\sqrt{6}$        &  $\Psi_{d7}^1$ \\ 
    & $-\frac{1}{2}$     &  $E_{1/2}$  &   $A_2 \otimes E_{1/2}$     &    $  \ket{|\bar{u}\bar{x}y+\bar{u}x\bar{y}-2u\bar{x}\bar{y}}/\sqrt{6}$     &  $\Psi_{d7}^2$ \\  
    \hline
\multirow{4}{2em} { $ue^2$ }         & \multirow{3}{2em}{$\pm\frac{1}{2}$}     &  $1E_{3/2}$  &  \multirow{3}{4em} { $E \otimes E_{1/2}$}     &    $  \ket{|(ux\bar{y}-u\bar{x}y)-i(\bar{u}\bar{x}y-\bar{u}x\bar{y})+i(ux\bar{x}-uy\bar{y})-(\bar{u}\bar{x}x-\bar{u}\bar{y}y)     }/2\sqrt{2}$     &  $\Psi_{d8}^1$ \\  
    &      &  $2E_{3/2}$  &        &    $  \ket{|(ux\bar{y}-u\bar{x}y)+i(\bar{u}\bar{x}y-\bar{u}x\bar{y})+i(ux\bar{x}-uy\bar{y})+(\bar{u}\bar{x}x-\bar{u}\bar{y}y)     }/2\sqrt{2}$    &  $\Psi_{d8}^2$ \\  
    &      &  $E_{1/2}$  &         &    $  \ket{|(ux\bar{y}-u\bar{x}y)-i(ux\bar{x}-uy\bar{y})}/2$     &  $\Psi_{d8}^3$ \\  
    &      &  $E_{1/2}$  &        &    $  \ket{|(\bar{u}\bar{x}y-\bar{u}x\bar{y})+i(\bar{u}\bar{x}x-\bar{u}\bar{y}y) }/2$     &  $\Psi_{d7}^4$ \\  
    \hline
\multirow{1}{2em} { $ue^2$ }    &    +$\frac{1}{2}$          &$E_{1/2}$  & $A_1 \otimes E_{1/2}$ &   $\ket{|ux\bar{x}+uy\bar{y} }/\sqrt{2}$   &   $\Psi_{d9}^1$ \\ 
     & -$\frac{1}{2}$      &  $E_{1/2}$  & $A_1 \otimes E_{1/2}$ &  $\ket{| \bar{u}\bar{x}x+\bar{u}\bar{y}y }/\sqrt{2}$    & $\Psi_{d9}^2$  \\
\hline  
  \end{tabular} 
\end{ruledtabular}
\end{table}

\end{widetext}

\subsubsection{Clebsh-Gordan expansion and Wigner-Eckart theorem}

For direct product of representations of a given group, the Clebsh-Gordan expansion  indicates how to make the decomposition. Accordingly, the direct product symmetry operator transforms the basis as  \cite{LudwigBook}: 
\begin{widetext}
\begin{equation}
\label{eq:projTP}
\begin{aligned}
P_R^{(\alpha \times \beta)} e^{(\alpha)}_i e^{(\beta)}_k :=&P_R e^{(\alpha)}_i  \otimes P_R e^{(\beta)}_k =\sum_{jl} D^{(\alpha)}_{ji}(R) D^{(\beta)}_{lk}(R)e^{(\alpha)}_j e^{(\beta)}_l = \sum_{jl} D^{(\alpha \times \beta )}_{jl,ik}(R) e^{(\alpha)}_j  e^{(\beta)}_l,  \\
\end{aligned}
\end{equation}
\end{widetext}
where the basis is
\begin{equation}
\{ e^{(\alpha \beta)}_{ij}  \} =  \{  e^{(\alpha)}_i e^{(\beta)}_j  | \text{ where } i=1,....,d_{\alpha}; j=1,....,d_{\beta}  \}
\end{equation}
If $D^{(\alpha)}$ and $D^{(\beta)}$ are irreducible representations, then  $D^{(\alpha \times \beta)}$ is in general a reducible representation.  The Clebsh-Gordan expansion  gives the decomposition detail from reducible representations to irreducible ones. If we define  $(\alpha \beta | \gamma)$  as  the Clebsh-Gordan coefficient (CGC) or reduction coefficient, the CGCs can be determined by : 
\begin{widetext}
 \begin{equation}
 \label{eq:GCC}
 \begin{aligned}
\sum_s \left( \begin{array}{cccc} \alpha & \beta & | &\gamma ,s   \\ i & k & |  & m \end{array}\right)  \left( \begin{array}{cccc} \alpha & \beta & | &\gamma ,s   \\ j & l & |  & n \end{array}\right)^*=\frac{d_{\gamma}}{g}\sum_R D^{(\alpha)}_{ij} (R) D^{(\beta)}_{kl}(R) D^{(\gamma)}_{mn} (R)^*
\end{aligned}
\end{equation}
\end{widetext}

Solving the above equation gives  the CGC table  for   ${C}_{{3v}}$, which are listed in Table \ref{table:CGC}. The results here are consistent with previous results \cite{MansonPRB2006,AltmannBook}. 

\begin{table}
\caption{Clebsch-Gordan coefficients of ${C}_{{3v}}$ irreducible representations in Cartesian coordinates.}
\label{table:CGC}
\begin{ruledtabular}
\scalebox{0.90}{
\begin{tabular}{ |c  c | } 
 \hline
   & \\
  $\left( \begin{array}{cccc} A_1 & A_1 & | &A_1  \\ 1 & 1 & |  & 1 \end{array}\right)=1$ $\quad$	&	$\left( \begin{array}{cccc} E & E & | &A_1  \\ j & k & |  & 1 \end{array}\right)= \frac{1}{\sqrt{2}} \left[ \begin{array}{cc} 1 & 0  \\ 0 & 1 \end{array}\right] $    \\
  $\left( \begin{array}{cccc} A_1 & A_2 & | &A_2  \\ 1 & 1 & |  & 1 \end{array}\right)=1$ $\quad$	&	$\left( \begin{array}{cccc} E & E & | &A_2  \\ j & k & |  & 1 \end{array}\right)= \frac{1}{\sqrt{2}} \left[ \begin{array}{cc} 0 & 1  \\ -1 & 0 \end{array}\right] $    \\
  $\left( \begin{array}{cccc} A_2 & A_2 & | &A_1  \\ 1 & 1 & |  & 1 \end{array}\right)=1$ $\quad$	&	 \\
  $\left( \begin{array}{cccc} A_1 & E & | &E  \\ 1 & j & |  & k \end{array}\right)=\left[ \begin{array}{cc} 1 & 0  \\ 0 & 1 \end{array}\right]$ $\quad$	&	$\left( \begin{array}{cccc} A_2 & E & | &E  \\ 1 & j & |  & k \end{array}\right)= \left[ \begin{array}{cc} 0 & 1  \\ -1 & 0 \end{array}\right]$    \\  
  & \\
  \hline 
\end{tabular} }
\end{ruledtabular}
\end{table}

The Winger-Eckart theorem\cite{CornwellBook} decomposes  the results of the operator on states of IRs with specific sub-indices as  the product  of the Clebsch-Gordan coefficient and a reduced  matrix elements depending only on the IR type:
\begin{equation}
\label{eq:wigner-eckart}
\bra{\psi^{\Gamma_f}_{k'}} O^{\Gamma_o}_p \ket{\psi^{\Gamma_i}_{k}}= \left( \begin{array}{cccc} \Gamma_i & \Gamma_o & | &\Gamma_f  \\ k & p & |  & k' \end{array}\right)^* \bra{\psi^{\Gamma_f}} |O^{\Gamma_o}| \ket{\psi^{\Gamma_i}}
\end{equation}
As we have included a systematic way to calculate the CGCs, many matrix elements can be simplified as the contraction term on the right in the above equation and the ratio among  matrix elements  of the same operator within the same IR types can be determined explicitly.  
\subsubsection{Selection Rules}
Selection rules state that for the general operator $O^{\prime}$ with symmetry type $\Gamma^{\prime}$ and states  $\ket{i}$ and $\ket{f}$  with symmetry type $\Gamma^{(f)}$ and $\Gamma^{(i)}$ respectively:
\begin{equation}
\label{eq:selection-rule}
 \Gamma^{\prime}  \otimes  \Gamma^{(f)} \not\supset \Gamma^{(i)} \Longrightarrow      \bra{i}O^{\prime}\ket{f}	\equiv 0  .
\end{equation}
The selection rule for an electric field among $C_{3v}$ group states are listed in Table \ref{table:selection_rule} .

\begin{table}
\caption{Optical transitions between multiplets in the  $C_{3v}$ symmetry group.}
\label{table:selection_rule}
\begin{center}
\begin{ruledtabular}
\begin{tabular}{ |c |c| c| c  | } 
 \hline
 $\Delta S=0 $  & $A_1$   & $A_2$  &  $ E$   \\
 \hline 
 $A_1$ & $\parallel $    &  0   &   $\perp $       \\ 
 $A_2$ &       & $\parallel $   &     $\perp $      \\ 
 $E$ 	    &      &     &    $\perp, \parallel $       \\ 
  \hline 
\end{tabular}\\
\end{ruledtabular}
\end{center}
\end{table}

\section{Phonons in ISC}
For ${C}_{{3v}}$ symmetry, phonon modes have  two IRs :  $A_1$ and $E$, and the strain tensor ($\epsilon_{ij}=\frac{\delta u_i}{\delta x_j}$) transforms as the linear basis product $x_i x_j$.  We can target on specific IRs and use the CGCs to explore how strain affects the system. We can get the strain Hamiltonian as the combination of projectors on single orbitals, i.e., Eq. \eqref{eq:Hstrain-new}.  To understand how the phonon modes affect the orbitals we first construct the  strain Hamiltonian  with respect to the manifold encompassing all  single orbitals of interest $\{e_x,e_y,u,v\}$: 
\begin{widetext}
\begin{equation}
\label{eq:Hstrain}
\begin{aligned}
H_{\text{strain}}= \delta^a_{A_1}A^a_1+\delta^b_{A_1}(A^b_1+A'^b_1+A''^b_1)+\delta^a_{E_1}E^a_1+\delta^a_{E_2}E^a_2+\delta^b_{E_1}(E^b_1+E'^b_1 )+\delta^b_{E_2}(E^b_2 +E'^b_2 ),
\end{aligned}
\end{equation}
\end{widetext}
where, $\delta^a_{A_1}=(e_{xx}+e_{yy})/2, \delta^b_{A_1}=e_{zz},\delta^a_{E_1}=(e_{xx}-e_{yy})/2,\delta^a_{E_2}= (e_{xy}+e_{yx})/2,\delta^b_{E_1}=(e_{xz}+e_{zx})/2,\delta^b_{E_2}=(e_{yz}+e_{zy})/2 $. The $z$ direction corresponds to $A_1$ IR according to which both $u$ and $v$ orbitals transform and the $A^a_1,A^b_1,A'^b_1A''^b_1,E^a_{1,2},E'^b_{1,2}$ are projectors on the single orbitals \cite{MazeNJP2011} in the basis of $\{e_x,e_y,u,v\}$ and are list below:
\begin{equation}
\label{eq:strain-proj}
\begin{aligned}
&A^a_1=\left( \begin{array}{cccc } 1& 0& 0  &0 \\ 0& 1& 0 &0   \\  0 & 0 & 0 & 0 \\ 0 & 0 & 0 & 0  \end{array} \right )  E^a_1=\left( \begin{array}{cccc } 1& 0& 0  &0 \\ 0& -1& 0 &0   \\  0 & 0 & 0 & 0 \\ 0 & 0 & 0 & 0  \end{array} \right )  E^a_2=\left( \begin{array}{cccc } 0& 1& 0  &0 \\ 1& 0& 0 &0   \\  0 & 0 & 0 & 0 \\ 0 & 0 & 0 & 0  \end{array} \right )  \\
&A^b_1=\left( \begin{array}{cccc } 0& 0& 0  &0 \\ 0& 0& 0 &0   \\  0 & 0 & 1 & 0 \\ 0 & 0 & 0 & 0  \end{array} \right )  E^b_1=\left( \begin{array}{cccc } 0& 0& 1  &0 \\ 0& 0& 0 &0   \\  1 & 0 & 0 & 0 \\ 0 & 0 & 0 & 0  \end{array} \right )  E^b_2=\left( \begin{array}{cccc } 0& 0& 0  &0 \\ 0& 0& 1 &0   \\  0 & 1 & 0 & 0 \\ 0 & 0 & 0 & 0  \end{array} \right )  \\
&A'^b_1=\left( \begin{array}{cccc } 0& 0& 0  &0 \\ 0& 0& 0 &0   \\  0 & 0 & 0 & 0 \\ 0 & 0 & 0 & 1  \end{array} \right )  E'^b_1=\left( \begin{array}{cccc } 0& 0& 0  &1 \\ 0& 0& 0 &0   \\  0 & 0 & 0 & 0 \\ 1 & 0 & 0 & 0  \end{array} \right )  E'^b_2=\left( \begin{array}{cccc } 0& 0& 0  &0 \\ 0& 0 & 0 &1   \\  0 & 0 & 0 & 0 \\ 0 & 1 & 0 & 0  \end{array} \right )  \\
&A''^b_1=\left( \begin{array}{cccc } 0& 0& 0  &0 \\ 0& 0& 0 &0   \\  0 & 0 & 0 & 1 \\ 0 & 0 & 1& 0  \end{array} \right ).
\end{aligned}
\end{equation} 

Reordering all terms to get a succinct projector: 
\begin{widetext}
\begin{equation}
\label{eq:Hstrain-new}
\begin{aligned}
H_{strain} & = \delta^a_{A_1} (\ket{x}\bra{x}+\ket{y}\bra{y}  ) +\delta^b_{A_1} (\ket{u}\bra{u}+\ket{u}\bra{v}+\ket{v}\bra{u}+\ket{v}\bra{v}  ) + \delta^a_{E_1} (\ket{x}\bra{x}-\ket{y}\bra{y}  )\\
& +\delta^b_{E_1} (\ket{x}\bra{u}+\ket{x}\bra{v}+\ket{u}\bra{x}+\ket{v}\bra{x}  ) + \delta^a_{E_2} ( \ket{x}\bra{y}+\ket{y}\bra{x} ) +\delta^b_{E_2} (\ket{y}\bra{u}+\ket{y}\bra{v}+\ket{u}\bra{y}+\ket{v}\bra{y}  ). \\
\end{aligned}
\end{equation}
\end{widetext}
The interaction of  phonons among 3-hole wave functions  can be constructed by using  Eq. \eqref{eq:Hamiltonian} and the projection rule for single orbitals. 
In the main text, we express the $\Gamma^{(1)}$ with the assumption that the quartets are in a ground vibrational mode, so the Eq. \eqref{eq:Gamma1} is an approximation.  The general version of the first order ISC is:
\begin{equation}
\label{eq:gamma-1-general}
\Gamma^{(1)} \propto\hbar |\lambda_{\bot(1,2)}|^2\sum_{nm} |\braket{\chi'_{\nu_m}}{\chi'_{\nu_n}}|^2\delta({\nu_n - \nu_m-\Delta}),   
\end{equation}
where the $\ket{\chi'_{\nu_{m}}}$,$\ket{\chi'_{\nu_{n}}}$ represent the general vibrational levels for quartet and target doublet respectively. 

The derivation of the second order ISC formula, Eq. \eqref{eq:Gamma2}, is as follows:
\begin{widetext}
\begin{eqnarray}
\Gamma_{2nd}  &=&  \frac{2\pi}{\hbar} \sum_{f,i} \left | \sum_{m}\frac{ \bra{f}V\ket{m}\bra{m}V\ket{i}  }{E_i-E_m} \right | ^2\delta(E_f-E_i)  \nonumber\\
&=& \frac{2\pi}{\hbar}\sum_{m,l,p,q,\pm1} \left | \frac{\bra{\Psi_{d_4},\chi_l}H_{eq}\ket{\Psi_{d_6},\chi_n}\bra{\Psi_{d_6},\chi_n}H_{soc}\ket{ \Psi_{{q1}^3_2} ,\chi_m   } }{E_{m}-E_{n}} \right |^2 \delta(E_{l}-E_{m})    .
\end{eqnarray}
The matrix elements of $H_{soc}$  are obtained from Table \ref{table:SOC}, and by using Eqs. \ref{eq:Gamma1} and \ref{eq:Hamiltonian}. Using the symbol $\alpha$ for the overall (unknown) numerical coefficient we have:
\begin{eqnarray}
\Gamma^{(2)}  &=& \alpha  \sum_{m,l,p,q,\pm1} \left | \sum_n  \frac{\braket{\chi_n}{\chi_m}\bra{\chi_l}   {\delta}_{pk} (a^\dagger_{p,k}+a_{p,k}) \ket{\chi_n}}{E_m-E_n} \right |^2 \delta(E_l-E_m) .
\end{eqnarray}
 Defining the electronic energy difference $E_{q1,\chi_0}-E_{d6,\chi_0}\equiv\Delta_6, E_{q1,\chi_0}-E_{d4,\chi_0}\equiv\Delta_4$ and using  $a^\dagger\ket{\chi_n}=\sqrt{n_{pq}+1}\ket{\chi^\dagger_n}$  and $a\ket{\chi_n}=\sqrt{n_{pq}}\ket{\chi^-_n}$ we obtain
\begin{eqnarray}
\Gamma^{(2)}  &=& \alpha  |\lambda_{\bot2}|^2 \sum_{m,l,p,q} \left[ \left | \sum_{n}  \frac{{\delta}_{pk} \braket{\chi_n}{\chi_m}\sqrt{n_{p,q}+1}\braket{\chi_l}{\chi^+_n}  }{E_m-E_n} \right |^2\delta(E_l-E_m)  + \left |\sum_n \frac{{\delta}_{pk} \braket{\chi_{n}}t{\chi_{m}}  \sqrt{n_{p,q}}\braket{\chi_l}{\chi^-_n}     }{E_m-E_{n}} \right |^2   \delta(E_l-E_m) \right ]  \nonumber  \\
&=& \alpha |\lambda_{\bot2}|^2 \sum_{m,l,p,q} \left [ \left | \sum_{n}  \frac{{\delta}_{pk} \braket{\chi_n}{\chi_m}\sqrt{n_{p,q}+1}\braket{\chi_l}{\chi^+_n}  }{\Delta_6+\nu_m-\nu_n-\omega_{p,q}} \right |^2\delta( \Delta_4+\nu_m-\nu_n-\omega_{p,q} ) \right . \nonumber \\
&+&  \left . \left | \sum_n \frac{{\delta}_{pk} \braket{\chi_{n}}{\chi_{m}}  \sqrt{n_{p,q}}\braket{\chi_l}{\chi^-_n}     }{\Delta_6+\nu_m-\nu_n+\omega_{p,q}} \right |^2  \right ] \delta(\Delta_4+\nu_m-\nu_n+\omega_{p,q}).
\end{eqnarray}
\end{widetext}
Other symbols represent the same as in Eq. \eqref{eq:Gamma2}. The general formula of $\Gamma^{(2)}$ includes the simple case especially if, e.g., the intermediate state is limited to just one phonon mode, $\chi_0$:
\begin{widetext}
\begin{eqnarray}
\Gamma^{(2)} &=& \alpha  |\lambda_{\bot2}|^2 \sum_{m,l,p,q}  | \tilde{\delta}_{pk}|^2|\braket{\chi_0}{\chi_m}  |^2 \left [    \frac{  (n_{p,q}+1)|\braket{\chi_l}{\chi^+_0} |^2 }{(\Delta_6+\nu_m-\omega_{p,q})^2} \delta( \Delta_4+\nu_m-\omega_{p,q} ) \right . \\
& +& \left . \frac{  (n_{p,q})|\braket{\chi_l}{\chi^-_0} |^2 }{(\Delta_6+\nu_m-\nu_n+\omega_{p,q})^2} \delta( \Delta_4+\nu_m-\nu_n+\omega_{p,q} )      \right]  \\
\end{eqnarray}
\end{widetext}
where the denominators reduce to $\omega^2_{pk}$  if we limit the $q2$ state vibration as $\chi_0$ only and the above equation simplifies as the version  in \cite{GoldmanPRB2015}. 

\section{Lindblad Terms}
For the first spin polarization protocol involving $g$, $q1$ quartets and $d4$, $d6$ doublets, we list the ISC Lindbladians:
\begin{equation}
\begin{aligned}
\label{lindbladians-ISC}
L(q1|{3}/{2}|,d6)&=\sqrt{\frac{3}{4}}\ket{d6}\bra{q1|{3}/{2}|}\sqrt{\gamma_{\text{ISC}}}  \\
L(q1|{1}/{2}|,d9)&=\sqrt{\frac{8}{3}}\ket{d9}\bra{q1|{1}/{2}|}\sqrt{\gamma_{\text{ISC}}}  \\
L(d4,g|{1}/{2}|)&=\sqrt{\frac{8}{3}}\ket{g|1/2|}\bra{d4}\sqrt{\gamma_{\text{ISC}}}  \\
L(q1|{1}/{2}|,d6)&=L(q1|{3}/{2}|,d9)=L(d4,g|{3}/{2}|) =0
\end{aligned}
\end{equation}
where   we choose ISC among electronic energy close wave functions but not the ones with large energy separation, in order  to have strong $\gamma_{\text{ISC}}$. The relaxation Lindlbadians (which could include possible photon and phonon relaxation) are: 
\begin{equation}
\begin{aligned}
\label{lindbladians-vertical}
L(g|{3}/{2}|,q1|{3}/{2}|)&=\ket{g|{3}/{2}|}\bra{q1|{3}/{2}|}\sqrt{\gamma_{{0}}} \\
L(g|{1}/{2}|,q1|{1}/{2}|)&=\ket{g|{1}/{2}|}\bra{q1|{1}/{2}|}\sqrt{\gamma_{{0}}} \\
L(d4,d6)&=\ket{d4}\bra{d6}\sqrt{\gamma_{{E}}}  \\
L(d4,d9)&=\ket{d4}\bra{d9}\sqrt{\gamma_{{A_1}}} . 
\end{aligned}
\end{equation}
We assume $\gamma_0=\gamma_{E}=\gamma_{A_1}$ in our calculation by treating them as a fast relaxation process.

For the second spin polarization channel, the corresponding ISC Lindbladians are: 
\begin{equation}
\begin{aligned}
\label{lindbladians2-ISC}
L(q2|{3}/{2}|,d2)&= c_1 \ket{d2}\bra{q2|{3}/{2}|}\sqrt{\gamma_{\text{ISC}}} \\
L(q2|{1}/{2}|,d2)&=c_1 \ket{d2}\bra{q2|{1}/{2}|}\sqrt{\gamma_{\text{ISC}}} \\
L(q2|{3}/{2}|,d3)&= c_1 \ket{d3}\bra{q2|{3}/{2}|}\sqrt{\gamma_{\text{ISC}}} \\
L(q2|{1}/{2}|,d3)&=c_1 \ket{d3}\bra{q2|{1}/{2}|}\sqrt{\gamma_{\text{ISC}}}, \\
\end{aligned}
\end{equation}
\begin{equation}
\begin{aligned}
L(q2|{3}/{2}|,d4)&= c_2 \ket{d4}\bra{q2|{3}/{2}|}\sqrt{\gamma_{\text{ISC}}} \\
L(q2|{1}/{2}|,d4)&=c_2 \ket{d4}\bra{q2|{1}/{2}|}\sqrt{\gamma_{\text{ISC}}}, \\
\end{aligned}
\end{equation}
where we change the ratio between $c_1$ and $c_2$ (hence different population preference among $d2$, $d3$, and $d4$) to have different spin polarization results (shown in Fig.  \ref{lindblad_q2});   and 
\begin{eqnarray}
L(d2,g|{3}/{2}|)&=&2\ket{g|3/2|}\bra{d2}\sqrt{\gamma_{\text{ISC}}} \nonumber \\
L(d2,g|{1}/{2}|)&=&2\sqrt{\frac{1}{3}}\ket{g|1/2|}\bra{d2}\sqrt{\gamma_{\text{ISC}}} \nonumber \\
L(d3,g|{3}/{2}|)&=&2\ket{g|3/2|}\bra{d3}\sqrt{\gamma_{\text{ISC}}} \nonumber \\
L(d3,g|{1}/{2}|)&=&2\sqrt{\frac{1}{3}}\ket{g|1/2|}\bra{d3}\sqrt{\gamma_{\text{ISC}}}. \nonumber \\
\end{eqnarray}
The relaxation Lindbladians are: 
\begin{eqnarray}
L(g|{3}/{2}|,q2|{3}/{2}|)&=\ket{g|{3}/{2}|}\bra{q1|{3}/{2}|}\sqrt{\gamma_{{0}}} \nonumber \\
L(g|{1}/{2}|,q2|{1}/{2}|)&=\ket{g|{1}/{2}|}\bra{q1|{1}/{2}|}\sqrt{\gamma_{{0}}} 
\end{eqnarray}
The $\gamma_0$ here is taken to be the same as the one of the first spin polarization channel.

\end{document}